\documentclass{article}
\usepackage{tikz}
\usepackage{enumerate}
\usepackage{subfigure}
\usepackage{hyperref}
\usepackage{enumitem}
\usepackage[a4paper,includeheadfoot,margin=2.54cm]{geometry}

\begin{document}
%

\title{Resource Selection for Federated Search on the Web}
\newcommand{\specialcell}[2][c]{%
  \begin{tabular}[#1]{@{}c@{}}#2\end{tabular}}
%
%
%
%
%


%
\author{
%
%
Dong Nguyen\textsuperscript {1}, Thomas Demeester\textsuperscript {2}, Dolf Trieschnigg\textsuperscript {1,3}, Djoerd Hiemstra\textsuperscript {1} \\
       {\textsuperscript {1} University of Twente, The Netherlands}\\
       {\textsuperscript {2} iMinds, Ghent University, Belgium}\\
       {\textsuperscript {3} MyDataFactory, The Netherlands}\\
       {\normalsize \{d.nguyen, r.b.trieschnigg, d.hiemstra\}@utwente.nl, tdmeeste@intec.ugent.be}
       }


\date{}

\maketitle

\begin{abstract}
A publicly available dataset for federated search reflecting a real web environment
has long been absent, making it difficult for researchers to test the validity of their federated search algorithms for the web setting. 
We present several experiments and analyses on resource selection on the web  using a recently released test collection containing the results
from more than a  hundred real search engines, ranging from large general web search engines such as Google, Bing and Yahoo
to small domain-specific engines. 
First, we experiment with estimating the size of uncooperative search engines on the web using query based sampling
and propose a new method using the ClueWeb09 dataset.  We find  the size estimates to be highly effective in resource selection. 
Second, we show that an optimized federated search system based on smaller web search engines can be an alternative to a system using large web search engines.
Third, we provide an empirical comparison of several popular resource selection methods and 
find that these methods are not readily suitable for resource selection on the web.
 Challenges include the sparse resource descriptions and extremely skewed
sizes of collections. 
\end{abstract}



\section{Introduction}

Despite the obvious utility and necessity of using large general web search engines for navigating the internet, a large part of the web cannot be reached by these. The so-called \emph{hidden} or \emph{deep web}, is not easily accessible  to web crawlers \cite{ilprints725}, and hence a large amount of content cannot be included in the web search indices.
Usually these pages are dynamic and only accessible through a specific search engine.
\emph{Distributed information retrieval}, also called  \emph{federated search}, provides a solution to this problem. 
The queries are directly issued to search interfaces of collections so that
crawling of these collections is not needed anymore.
In this paper we focus on one specific task in federated search: \emph{resource selection}, which aims at selecting suitable collections to forward a specific query to.

A large amount of research has been done in the field of federated search. However, an appropriate dataset reflecting
a realistic web environment remained absent until recently. 
As a result, many proposed resource selection methods have been evaluated 
on distributed collections that were artificially composed by dividing TREC collections \cite{DBLP:journals/ftir/ShokouhiS11}, according to several criteria (like topic or source). These test collections are very different from real search engines we find on the web, which have
different retrieval methods, skewed sizes and heterogeneous content types (images, text, video etc).
In addition, some of the proposed methods for resource selection make strong assumptions that may not hold on the web.
For example, some of the proposed `big document models' assume that all resources use
a certain ranking method, while the `small document models' are built on the idea that
we are able to obtain a good ranking on a centralized sample index. 
As a result, it is not clear to what extent the proposed resource selection methods are suited for the web. 

In this paper we use a newly released dataset presented in Nguyen et al. \cite{nguyencikm2012}.
The dataset contains result pages from 108 actual web search engines, ranging from large general web search engines such as Google, Bing and Yahoo to small domain-specific engines.

We propose a new method for the estimation of collection sizes, based on samples obtained by query-based sampling, and assuming the ClueWeb09 web collection is a representative sample from the web. 
We also investigate the feasibility of web search without the use of general web search engines. For that, we compare an optimized federated search system based on smaller web search engines, with an optimized system using the large general web search engines.
Finally, we present an empirical comparison of well-known resource selection methods, 
finding that they are not entirely suitable for this setting.
 
The remaining part of the paper is structured as follows. 
Section 2 discusses related work. Then, the dataset is described.
In section 4, we present experiments on size estimation.
In section 5
several benchmarking experiments are presented. We conclude with a discussion and future work.

\section{Related work}
A federated search system presents three challenges: \emph{resource description}, obtaining a 
representation of the resource, \emph{resource selection}, selecting suitable
collections for a query and \emph{resource merging}, creating a single ranked list from the returned results \cite{callan2000}.
In this section we will focus on relevant work related to resource selection. We will also briefly 
review research on estimating the size of collections.

\subsection{Resource selection}
During the resource selection step, the broker selects a subset of the collections to send the query to.
Several approaches have been explored for the resource selection problem. 
One line of research involves the \emph{big document} models. These models represent each collection as one big document.
Standard retrieval methods (or adaptations of them) are used to rank the big documents (collections) for a query.
Examples are CORI \cite{Callan95searchingdistributed}, which uses a modified version of INQUERY, 
and language modeling approaches \cite{Si:2002:LMF:584792.584856,Xu:1999:CLM:312624.312687}.
Losing the document boundaries has several disadvantages.
For example, the contents of the big document can be dominated by a few large documents.

To retain the document boundaries, \emph{small document models} have been proposed. 
These have been found to be more effective than big document models.
A query is issued on a centralized index of the document samples. The proposed methods
vary in how the ranking or scores are used to select the collections.
ReDDE \cite{Si:2003:RDD:860435.860490} and CRCS \cite{Shokouhi:2007:CCS:1763653.1763674}
use the ranking of the documents in the sample index and a scaling factor to handle the varying sizes of collections. 
Contrary to ReDDE, the weight of a document in CRCS depends on the rank, while ReDDE uses only a cut off.
GAVG \cite{Seo:2008:BSS:1458082.1458222} uses the score of the documents instead of the ranking,
by calculating the geometric mean of the query likelihood of the top documents for each collection.
These methods assume that the system is able to obtain a good ranking on a centralized sample index.
However, when having multiple media types (e.g.  images and text), this assumption may not hold.

Recently, methods have been proposed that frame the collection selection problem as a \emph{classification} problem
(e.g. \cite{Arguello:2009:CRS:1645953.1646115}, \cite{Hong:2010:JPC:1835449.1835468} and  \cite{Kim:2010:RUM:1835449.1835461}).
The main benefit of this approach is the ability to easily add different types of evidence for prediction and not only
rely on content-based methods. For example, Arguello et al. \cite{Arguello:2009:CRS:1645953.1646115}
used logistic regression models and explored corpus features (such as CORI and ReDDE), 
query category features and click-through features.
Hong et al. \cite{Hong:2010:JPC:1835449.1835468} also used logistic classification models, but added a term to account for the similarity between resources.
This was motivated by the idea that a resource tends to be more relevant when similar resources
have a high probability of being relevant. 

Craswell et al. \cite{Craswell:2000:SSW:336597.336628} experimented with resource selection for the web. 
To simulate a web environment where search engines have different retrieval algorithms, 
resources used different retrieval methods such as BM25 and Boolean ranking.
Hawking and Thomas \cite{Hawking:2005:SSM:1076034.1076050}  evaluated resource selection methods in the GOV domain 
and proposed a hybrid approach combining distributed and central IR techniques.
Although the work just discussed used actual web pages, the collections were relatively small compared to collections found
on the web. For example, the largest collections were around 20 thousand documents \cite{Hawking:2005:SSM:1076034.1076050},
or a couple of thousand \cite{Craswell:2000:SSW:336597.336628}.

\subsection{Size estimation}
Many of the resource selection methods incorporate the size of a collection
in their scoring method. However, when dealing with uncooperative resources,
the actual size is not known and has to be estimated using the obtained samples.
Methods to estimate the size of a collection assume random samples.
Unfortunately, the samples from query based sampling are often not random, resulting
in additional noise or challenges when estimating the size. 

The Capture-Recapture \cite{Liu:2002:DRS:584792.584909} method uses the overlap
of documents from two random samples to estimate the size, while
Capture History \cite{Shokouhi:2006:CCS:1148170.1148227} looks at overlapping
documents when having a consecutive number of random samples.
{Sample - Resample \cite{Si:2003:RDD:860435.860490}
relies on document frequencies. Since many uncooperative resources 
do not provide (accurate) document frequencies, this method is often not suitable.
Broder et al. \cite{Broder:2006:ECS:1183614.1183699} proposed two approaches to estimate the corpus size
via query pools. The first relies on a uniform random sample of documents.
The second approach is more practical and uses two (fairly) uncorrelated query pools 
and is related to the Capture-Recapture method.

\section{Dataset description}
In this section we briefly summarize the test collection presented in Nguyen et al.\ \cite{nguyencikm2012}.\\

\noindent \textbf{Motivation}\\
The dataset was developed to reflect a realistic web environment, and designed to have the following properties:
heterogeneous content, skewed sizes and relevance distributions, different retrieval algorithms
and overlapping documents.\\

\noindent \textbf{Search engines}\\
\noindent The dataset contains 108 search engines, with a large variation in size and topic (e.g. sport, general web, news,
academic, etc.).  The dataset contains general web search engines such as Google and Bing, as well as smaller search
engines focusing on a single domain or topic (for example arXiv.org, Photobucket,  Starbucks and TechRepublic). \\

\noindent \textbf{Sampling}\\
\noindent Using query based sampling, samples were collected to inform the resource selection methods. 
The snippets as well as the documents of the search results were stored.
Three methods were used to select queries for sampling:
\emph{Random},  \emph{Top}, and \emph{Zipf}. \emph{Random} 
selects single terms randomly from the documents sampled so far. \emph{Top}
uses the most popular queries from a web search engine's query log
\cite{Pass:2006:PS:1146847.1146848}. \emph{Zipf} uses single term queries taken evenly
from the binned term distribution in ClueWeb09.
For each query, the top 10 results from each engine were retrieved.
Sampling was done by collecting the snippets of 197 queries and
downloading the actual documents of the first 40 queries.
\\

\noindent \textbf{Topics}\\
\noindent For testing, the dataset contains results of the search engines for the 50 topics of TREC Web Track 2010. These queries were developed to represent web search queries and are therefore very short. In addition, they were used to evaluate the Web Track diversity track and therefore many of them are ambiguous.\\

\noindent \textbf{Relevance judgements}\\
\noindent For the set of test queries, the relevance judgements for the top 10 results of each search engine were collected.
Because many of the topics are ambiguous, annotators  were instructed to judge according to the general information need indicated 
in the TREC Web Track 2010 full topic specification. For example for the query \emph{titan}
the general information need is \emph{Find information on the Nissan Titan truck}.
For the experiments in this paper, we use the relevance judgements based on the pages, but snippet judgments were collected as well.
Six levels of page relevance were used, denoted in increasing order of relevance as \emph{Junk},  \emph{Non}, \emph{Rel}, \emph{HRel}, \emph{Key}, and \emph{Nav}, corresponding with the relevance levels from the Web TREC 2010 ad-hoc task~\cite{trec2010}. For our analysis, we merged the \emph{Junk} results into the \emph{Non} category.  
For most of the experiments in this paper, a binary relevance is used where pages are considered relevant when scored on average
better than \emph{Rel}.

\section{Size estimation}
Estimating the size of a collection is a key step in many resource selection methods.
In this section we discuss the used methods. In the next section we analyze the effect
of the size estimation methods on resource selection.
Note that the ratio of sizes among the collections is more important for resource selection experiments, than absolute numbers. 
We will reuse the samples obtained from query based sampling (to inform resource selection)
to estimate the collection sizes. We explore two different methods to estimate the size of a collection.
The first method involves a reference corpus (in our case ClueWeb09), the second method
makes use of the number of overlapping documents in the samples. \\

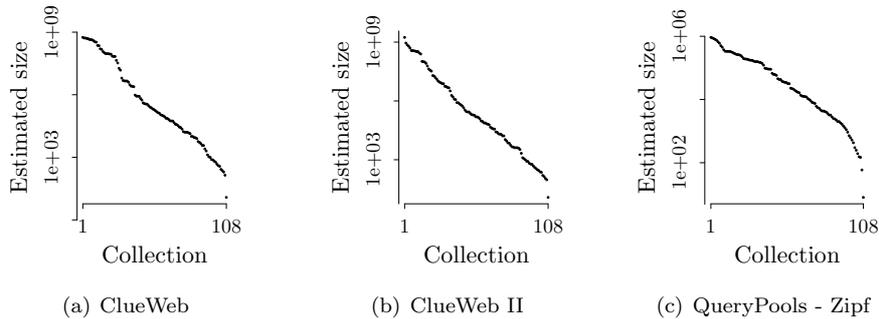
\begin{figure*}
\begin{center}
\mbox{
\subfigure[ClueWeb]{
\scalebox{0.30}{

\begin{tikzpicture}[x=1pt,y=1pt]
\definecolor[named]{drawColor}{rgb}{0.00,0.00,0.00}
\definecolor[named]{fillColor}{rgb}{1.00,1.00,1.00}
\fill[color=fillColor,] (0,0) rectangle (361.35,361.35);
\begin{scope}
\path[clip] (120.00, 96.00) rectangle (313.35,313.35);
\definecolor[named]{drawColor}{rgb}{0.00,0.00,0.00}
\definecolor[named]{fillColor}{rgb}{0.00,0.00,0.00}

\draw[color=drawColor,line cap=round,line join=round,fill=fillColor,] (127.16,305.30) circle (  1.12);

\draw[color=drawColor,line cap=round,line join=round,fill=fillColor,] (128.83,304.55) circle (  1.12);

\draw[color=drawColor,line cap=round,line join=round,fill=fillColor,] (130.51,304.44) circle (  1.12);

\draw[color=drawColor,line cap=round,line join=round,fill=fillColor,] (132.18,303.86) circle (  1.12);

\draw[color=drawColor,line cap=round,line join=round,fill=fillColor,] (133.85,303.63) circle (  1.12);

\draw[color=drawColor,line cap=round,line join=round,fill=fillColor,] (135.53,302.58) circle (  1.12);

\draw[color=drawColor,line cap=round,line join=round,fill=fillColor,] (137.20,302.49) circle (  1.12);

\draw[color=drawColor,line cap=round,line join=round,fill=fillColor,] (138.87,302.13) circle (  1.12);

\draw[color=drawColor,line cap=round,line join=round,fill=fillColor,] (140.55,301.10) circle (  1.12);

\draw[color=drawColor,line cap=round,line join=round,fill=fillColor,] (142.22,299.67) circle (  1.12);

\draw[color=drawColor,line cap=round,line join=round,fill=fillColor,] (143.89,299.30) circle (  1.12);

\draw[color=drawColor,line cap=round,line join=round,fill=fillColor,] (145.57,295.16) circle (  1.12);

\draw[color=drawColor,line cap=round,line join=round,fill=fillColor,] (147.24,294.80) circle (  1.12);

\draw[color=drawColor,line cap=round,line join=round,fill=fillColor,] (148.91,290.89) circle (  1.12);

\draw[color=drawColor,line cap=round,line join=round,fill=fillColor,] (150.59,290.21) circle (  1.12);

\draw[color=drawColor,line cap=round,line join=round,fill=fillColor,] (152.26,287.48) circle (  1.12);

\draw[color=drawColor,line cap=round,line join=round,fill=fillColor,] (153.93,285.98) circle (  1.12);

\draw[color=drawColor,line cap=round,line join=round,fill=fillColor,] (155.60,284.79) circle (  1.12);

\draw[color=drawColor,line cap=round,line join=round,fill=fillColor,] (157.28,284.77) circle (  1.12);

\draw[color=drawColor,line cap=round,line join=round,fill=fillColor,] (158.95,284.76) circle (  1.12);

\draw[color=drawColor,line cap=round,line join=round,fill=fillColor,] (160.62,284.35) circle (  1.12);

\draw[color=drawColor,line cap=round,line join=round,fill=fillColor,] (162.30,284.06) circle (  1.12);

\draw[color=drawColor,line cap=round,line join=round,fill=fillColor,] (163.97,281.61) circle (  1.12);

\draw[color=drawColor,line cap=round,line join=round,fill=fillColor,] (165.64,281.29) circle (  1.12);

\draw[color=drawColor,line cap=round,line join=round,fill=fillColor,] (167.32,281.29) circle (  1.12);

\draw[color=drawColor,line cap=round,line join=round,fill=fillColor,] (168.99,276.36) circle (  1.12);

\draw[color=drawColor,line cap=round,line join=round,fill=fillColor,] (170.66,273.11) circle (  1.12);

\draw[color=drawColor,line cap=round,line join=round,fill=fillColor,] (172.34,266.34) circle (  1.12);

\draw[color=drawColor,line cap=round,line join=round,fill=fillColor,] (174.01,264.06) circle (  1.12);

\draw[color=drawColor,line cap=round,line join=round,fill=fillColor,] (175.68,254.09) circle (  1.12);

\draw[color=drawColor,line cap=round,line join=round,fill=fillColor,] (177.36,251.07) circle (  1.12);

\draw[color=drawColor,line cap=round,line join=round,fill=fillColor,] (179.03,250.49) circle (  1.12);

\draw[color=drawColor,line cap=round,line join=round,fill=fillColor,] (180.70,250.44) circle (  1.12);

\draw[color=drawColor,line cap=round,line join=round,fill=fillColor,] (182.38,250.39) circle (  1.12);

\draw[color=drawColor,line cap=round,line join=round,fill=fillColor,] (184.05,248.95) circle (  1.12);

\draw[color=drawColor,line cap=round,line join=round,fill=fillColor,] (185.72,245.72) circle (  1.12);

\draw[color=drawColor,line cap=round,line join=round,fill=fillColor,] (187.39,243.93) circle (  1.12);

\draw[color=drawColor,line cap=round,line join=round,fill=fillColor,] (189.07,243.75) circle (  1.12);

\draw[color=drawColor,line cap=round,line join=round,fill=fillColor,] (190.74,243.22) circle (  1.12);

\draw[color=drawColor,line cap=round,line join=round,fill=fillColor,] (192.41,232.65) circle (  1.12);

\draw[color=drawColor,line cap=round,line join=round,fill=fillColor,] (194.09,231.35) circle (  1.12);

\draw[color=drawColor,line cap=round,line join=round,fill=fillColor,] (195.76,231.32) circle (  1.12);

\draw[color=drawColor,line cap=round,line join=round,fill=fillColor,] (197.43,231.22) circle (  1.12);

\draw[color=drawColor,line cap=round,line join=round,fill=fillColor,] (199.11,228.46) circle (  1.12);

\draw[color=drawColor,line cap=round,line join=round,fill=fillColor,] (200.78,226.42) circle (  1.12);

\draw[color=drawColor,line cap=round,line join=round,fill=fillColor,] (202.45,222.85) circle (  1.12);

\draw[color=drawColor,line cap=round,line join=round,fill=fillColor,] (204.13,221.91) circle (  1.12);

\draw[color=drawColor,line cap=round,line join=round,fill=fillColor,] (205.80,221.52) circle (  1.12);

\draw[color=drawColor,line cap=round,line join=round,fill=fillColor,] (207.47,221.24) circle (  1.12);

\draw[color=drawColor,line cap=round,line join=round,fill=fillColor,] (209.15,219.40) circle (  1.12);

\draw[color=drawColor,line cap=round,line join=round,fill=fillColor,] (210.82,217.60) circle (  1.12);

\draw[color=drawColor,line cap=round,line join=round,fill=fillColor,] (212.49,216.87) circle (  1.12);

\draw[color=drawColor,line cap=round,line join=round,fill=fillColor,] (214.17,215.28) circle (  1.12);

\draw[color=drawColor,line cap=round,line join=round,fill=fillColor,] (215.84,214.60) circle (  1.12);

\draw[color=drawColor,line cap=round,line join=round,fill=fillColor,] (217.51,213.20) circle (  1.12);

\draw[color=drawColor,line cap=round,line join=round,fill=fillColor,] (219.18,211.92) circle (  1.12);

\draw[color=drawColor,line cap=round,line join=round,fill=fillColor,] (220.86,210.58) circle (  1.12);

\draw[color=drawColor,line cap=round,line join=round,fill=fillColor,] (222.53,210.03) circle (  1.12);

\draw[color=drawColor,line cap=round,line join=round,fill=fillColor,] (224.20,208.00) circle (  1.12);

\draw[color=drawColor,line cap=round,line join=round,fill=fillColor,] (225.88,207.32) circle (  1.12);

\draw[color=drawColor,line cap=round,line join=round,fill=fillColor,] (227.55,206.57) circle (  1.12);

\draw[color=drawColor,line cap=round,line join=round,fill=fillColor,] (229.22,205.46) circle (  1.12);

\draw[color=drawColor,line cap=round,line join=round,fill=fillColor,] (230.90,203.57) circle (  1.12);

\draw[color=drawColor,line cap=round,line join=round,fill=fillColor,] (232.57,203.36) circle (  1.12);

\draw[color=drawColor,line cap=round,line join=round,fill=fillColor,] (234.24,202.05) circle (  1.12);

\draw[color=drawColor,line cap=round,line join=round,fill=fillColor,] (235.92,200.38) circle (  1.12);

\draw[color=drawColor,line cap=round,line join=round,fill=fillColor,] (237.59,200.22) circle (  1.12);

\draw[color=drawColor,line cap=round,line join=round,fill=fillColor,] (239.26,200.10) circle (  1.12);

\draw[color=drawColor,line cap=round,line join=round,fill=fillColor,] (240.94,198.13) circle (  1.12);

\draw[color=drawColor,line cap=round,line join=round,fill=fillColor,] (242.61,196.05) circle (  1.12);

\draw[color=drawColor,line cap=round,line join=round,fill=fillColor,] (244.28,195.63) circle (  1.12);

\draw[color=drawColor,line cap=round,line join=round,fill=fillColor,] (245.96,194.32) circle (  1.12);

\draw[color=drawColor,line cap=round,line join=round,fill=fillColor,] (247.63,192.44) circle (  1.12);

\draw[color=drawColor,line cap=round,line join=round,fill=fillColor,] (249.30,191.90) circle (  1.12);

\draw[color=drawColor,line cap=round,line join=round,fill=fillColor,] (250.97,189.90) circle (  1.12);

\draw[color=drawColor,line cap=round,line join=round,fill=fillColor,] (252.65,186.68) circle (  1.12);

\draw[color=drawColor,line cap=round,line join=round,fill=fillColor,] (254.32,186.06) circle (  1.12);

\draw[color=drawColor,line cap=round,line join=round,fill=fillColor,] (255.99,186.06) circle (  1.12);

\draw[color=drawColor,line cap=round,line join=round,fill=fillColor,] (257.67,185.90) circle (  1.12);

\draw[color=drawColor,line cap=round,line join=round,fill=fillColor,] (259.34,185.14) circle (  1.12);

\draw[color=drawColor,line cap=round,line join=round,fill=fillColor,] (261.01,184.22) circle (  1.12);

\draw[color=drawColor,line cap=round,line join=round,fill=fillColor,] (262.69,180.77) circle (  1.12);

\draw[color=drawColor,line cap=round,line join=round,fill=fillColor,] (264.36,180.53) circle (  1.12);

\draw[color=drawColor,line cap=round,line join=round,fill=fillColor,] (266.03,179.35) circle (  1.12);

\draw[color=drawColor,line cap=round,line join=round,fill=fillColor,] (267.71,179.11) circle (  1.12);

\draw[color=drawColor,line cap=round,line join=round,fill=fillColor,] (269.38,177.97) circle (  1.12);

\draw[color=drawColor,line cap=round,line join=round,fill=fillColor,] (271.05,174.28) circle (  1.12);

\draw[color=drawColor,line cap=round,line join=round,fill=fillColor,] (272.73,173.24) circle (  1.12);

\draw[color=drawColor,line cap=round,line join=round,fill=fillColor,] (274.40,171.74) circle (  1.12);

\draw[color=drawColor,line cap=round,line join=round,fill=fillColor,] (276.07,169.53) circle (  1.12);

\draw[color=drawColor,line cap=round,line join=round,fill=fillColor,] (277.75,169.32) circle (  1.12);

\draw[color=drawColor,line cap=round,line join=round,fill=fillColor,] (279.42,165.03) circle (  1.12);

\draw[color=drawColor,line cap=round,line join=round,fill=fillColor,] (281.09,162.43) circle (  1.12);

\draw[color=drawColor,line cap=round,line join=round,fill=fillColor,] (282.76,157.55) circle (  1.12);

\draw[color=drawColor,line cap=round,line join=round,fill=fillColor,] (284.44,155.15) circle (  1.12);

\draw[color=drawColor,line cap=round,line join=round,fill=fillColor,] (286.11,154.40) circle (  1.12);

\draw[color=drawColor,line cap=round,line join=round,fill=fillColor,] (287.78,151.60) circle (  1.12);

\draw[color=drawColor,line cap=round,line join=round,fill=fillColor,] (289.46,151.18) circle (  1.12);

\draw[color=drawColor,line cap=round,line join=round,fill=fillColor,] (291.13,149.62) circle (  1.12);

\draw[color=drawColor,line cap=round,line join=round,fill=fillColor,] (292.80,148.43) circle (  1.12);

\draw[color=drawColor,line cap=round,line join=round,fill=fillColor,] (294.48,145.75) circle (  1.12);

\draw[color=drawColor,line cap=round,line join=round,fill=fillColor,] (296.15,144.01) circle (  1.12);

\draw[color=drawColor,line cap=round,line join=round,fill=fillColor,] (297.82,143.15) circle (  1.12);

\draw[color=drawColor,line cap=round,line join=round,fill=fillColor,] (299.50,139.31) circle (  1.12);

\draw[color=drawColor,line cap=round,line join=round,fill=fillColor,] (301.17,137.80) circle (  1.12);

\draw[color=drawColor,line cap=round,line join=round,fill=fillColor,] (302.84,135.94) circle (  1.12);

\draw[color=drawColor,line cap=round,line join=round,fill=fillColor,] (304.52,131.96) circle (  1.12);

\draw[color=drawColor,line cap=round,line join=round,fill=fillColor,] (306.19,104.05) circle (  1.12);
\end{scope}
\begin{scope}
\path[clip] (  0.00,  0.00) rectangle (361.35,361.35);
\definecolor[named]{drawColor}{rgb}{0.00,0.00,0.00}

\node[color=drawColor,anchor=base,inner sep=0pt, outer sep=0pt, scale=  3.00] at (216.67, 21.60) {Collection%
};

\node[rotate= 90.00,color=drawColor,anchor=base,inner sep=0pt, outer sep=0pt, scale=  3.00] at ( 57.60,204.67) {Estimated size%
};
\end{scope}
\begin{scope}
\path[clip] (  0.00,  0.00) rectangle (361.35,361.35);
\definecolor[named]{drawColor}{rgb}{0.00,0.00,0.00}

\draw[color=drawColor,line cap=round,line join=round,fill opacity=0.00,] (120.00, 75.74) -- (120.00,311.87);

\draw[color=drawColor,line cap=round,line join=round,fill opacity=0.00,] (120.00, 75.74) -- (114.00, 75.74);

\draw[color=drawColor,line cap=round,line join=round,fill opacity=0.00,] (120.00,154.45) -- (114.00,154.45);

\draw[color=drawColor,line cap=round,line join=round,fill opacity=0.00,] (120.00,233.16) -- (114.00,233.16);

\draw[color=drawColor,line cap=round,line join=round,fill opacity=0.00,] (120.00,311.87) -- (114.00,311.87);

\node[rotate= 90.00,color=drawColor,anchor=base,inner sep=0pt, outer sep=0pt, scale=  2.50] at ( 96.00, 154.45) {1e+03%
};

\node[rotate= 90.00,color=drawColor,anchor=base,inner sep=0pt, outer sep=0pt, scale=  2.50] at ( 96.00,311.87) {1e+09%
};

\draw[color=drawColor,line cap=round,line join=round,fill opacity=0.00,] (127.16, 96.00) -- (306.19, 96.00);

\draw[color=drawColor,line cap=round,line join=round,fill opacity=0.00,] (127.16, 96.00) -- (127.16, 90.00);

\draw[color=drawColor,line cap=round,line join=round,fill opacity=0.00,] (306.19, 96.00) -- (306.19, 90.00);

\node[color=drawColor,anchor=base,inner sep=0pt, outer sep=0pt, scale=  2.50] at (127.16, 60.00) {1%
};

\node[color=drawColor,anchor=base,inner sep=0pt, outer sep=0pt, scale=  2.50] at (306.19, 60.00) {108%
};
\end{scope}
\end{tikzpicture}

}
}
\subfigure[ClueWeb II]{
\scalebox{0.30}{

\begin{tikzpicture}[x=1pt,y=1pt]
\definecolor[named]{drawColor}{rgb}{0.00,0.00,0.00}
\definecolor[named]{fillColor}{rgb}{1.00,1.00,1.00}
\fill[color=fillColor,fill opacity=0.00,] (0,0) rectangle (361.35,361.35);
\begin{scope}
\path[clip] (120.00, 96.00) rectangle (313.35,313.35);
\definecolor[named]{drawColor}{rgb}{0.00,0.00,0.00}
\definecolor[named]{fillColor}{rgb}{0.00,0.00,0.00}

\draw[color=drawColor,line cap=round,line join=round,fill=fillColor,] (127.16,305.30) circle (  1.12);

\draw[color=drawColor,line cap=round,line join=round,fill=fillColor,] (128.83,298.20) circle (  1.12);

\draw[color=drawColor,line cap=round,line join=round,fill=fillColor,] (130.51,295.74) circle (  1.12);

\draw[color=drawColor,line cap=round,line join=round,fill=fillColor,] (132.18,294.24) circle (  1.12);

\draw[color=drawColor,line cap=round,line join=round,fill=fillColor,] (133.85,291.98) circle (  1.12);

\draw[color=drawColor,line cap=round,line join=round,fill=fillColor,] (135.53,288.78) circle (  1.12);

\draw[color=drawColor,line cap=round,line join=round,fill=fillColor,] (137.20,288.53) circle (  1.12);

\draw[color=drawColor,line cap=round,line join=round,fill=fillColor,] (138.87,288.43) circle (  1.12);

\draw[color=drawColor,line cap=round,line join=round,fill=fillColor,] (140.55,287.96) circle (  1.12);

\draw[color=drawColor,line cap=round,line join=round,fill=fillColor,] (142.22,287.40) circle (  1.12);

\draw[color=drawColor,line cap=round,line join=round,fill=fillColor,] (143.89,287.40) circle (  1.12);

\draw[color=drawColor,line cap=round,line join=round,fill=fillColor,] (145.57,285.57) circle (  1.12);

\draw[color=drawColor,line cap=round,line join=round,fill=fillColor,] (147.24,284.37) circle (  1.12);

\draw[color=drawColor,line cap=round,line join=round,fill=fillColor,] (148.91,275.05) circle (  1.12);

\draw[color=drawColor,line cap=round,line join=round,fill=fillColor,] (150.59,274.68) circle (  1.12);

\draw[color=drawColor,line cap=round,line join=round,fill=fillColor,] (152.26,274.26) circle (  1.12);

\draw[color=drawColor,line cap=round,line join=round,fill=fillColor,] (153.93,272.57) circle (  1.12);

\draw[color=drawColor,line cap=round,line join=round,fill=fillColor,] (155.60,269.28) circle (  1.12);

\draw[color=drawColor,line cap=round,line join=round,fill=fillColor,] (157.28,265.54) circle (  1.12);

\draw[color=drawColor,line cap=round,line join=round,fill=fillColor,] (158.95,263.90) circle (  1.12);

\draw[color=drawColor,line cap=round,line join=round,fill=fillColor,] (160.62,262.86) circle (  1.12);

\draw[color=drawColor,line cap=round,line join=round,fill=fillColor,] (162.30,257.31) circle (  1.12);

\draw[color=drawColor,line cap=round,line join=round,fill=fillColor,] (163.97,256.61) circle (  1.12);

\draw[color=drawColor,line cap=round,line join=round,fill=fillColor,] (165.64,254.63) circle (  1.12);

\draw[color=drawColor,line cap=round,line join=round,fill=fillColor,] (167.32,252.66) circle (  1.12);

\draw[color=drawColor,line cap=round,line join=round,fill=fillColor,] (168.99,250.28) circle (  1.12);

\draw[color=drawColor,line cap=round,line join=round,fill=fillColor,] (170.66,249.09) circle (  1.12);

\draw[color=drawColor,line cap=round,line join=round,fill=fillColor,] (172.34,248.13) circle (  1.12);

\draw[color=drawColor,line cap=round,line join=round,fill=fillColor,] (174.01,247.19) circle (  1.12);

\draw[color=drawColor,line cap=round,line join=round,fill=fillColor,] (175.68,246.97) circle (  1.12);

\draw[color=drawColor,line cap=round,line join=round,fill=fillColor,] (177.36,243.75) circle (  1.12);

\draw[color=drawColor,line cap=round,line join=round,fill=fillColor,] (179.03,243.39) circle (  1.12);

\draw[color=drawColor,line cap=round,line join=round,fill=fillColor,] (180.70,242.30) circle (  1.12);

\draw[color=drawColor,line cap=round,line join=round,fill=fillColor,] (182.38,241.86) circle (  1.12);

\draw[color=drawColor,line cap=round,line join=round,fill=fillColor,] (184.05,237.95) circle (  1.12);

\draw[color=drawColor,line cap=round,line join=round,fill=fillColor,] (185.72,232.33) circle (  1.12);

\draw[color=drawColor,line cap=round,line join=round,fill=fillColor,] (187.39,230.46) circle (  1.12);

\draw[color=drawColor,line cap=round,line join=round,fill=fillColor,] (189.07,227.71) circle (  1.12);

\draw[color=drawColor,line cap=round,line join=round,fill=fillColor,] (190.74,226.99) circle (  1.12);

\draw[color=drawColor,line cap=round,line join=round,fill=fillColor,] (192.41,223.28) circle (  1.12);

\draw[color=drawColor,line cap=round,line join=round,fill=fillColor,] (194.09,223.07) circle (  1.12);

\draw[color=drawColor,line cap=round,line join=round,fill=fillColor,] (195.76,221.58) circle (  1.12);

\draw[color=drawColor,line cap=round,line join=round,fill=fillColor,] (197.43,220.34) circle (  1.12);

\draw[color=drawColor,line cap=round,line join=round,fill=fillColor,] (199.11,218.95) circle (  1.12);

\draw[color=drawColor,line cap=round,line join=round,fill=fillColor,] (200.78,216.75) circle (  1.12);

\draw[color=drawColor,line cap=round,line join=round,fill=fillColor,] (202.45,213.83) circle (  1.12);

\draw[color=drawColor,line cap=round,line join=round,fill=fillColor,] (204.13,211.95) circle (  1.12);

\draw[color=drawColor,line cap=round,line join=round,fill=fillColor,] (205.80,211.52) circle (  1.12);

\draw[color=drawColor,line cap=round,line join=round,fill=fillColor,] (207.47,211.45) circle (  1.12);

\draw[color=drawColor,line cap=round,line join=round,fill=fillColor,] (209.15,209.54) circle (  1.12);

\draw[color=drawColor,line cap=round,line join=round,fill=fillColor,] (210.82,208.07) circle (  1.12);

\draw[color=drawColor,line cap=round,line join=round,fill=fillColor,] (212.49,207.79) circle (  1.12);

\draw[color=drawColor,line cap=round,line join=round,fill=fillColor,] (214.17,207.30) circle (  1.12);

\draw[color=drawColor,line cap=round,line join=round,fill=fillColor,] (215.84,205.45) circle (  1.12);

\draw[color=drawColor,line cap=round,line join=round,fill=fillColor,] (217.51,205.31) circle (  1.12);

\draw[color=drawColor,line cap=round,line join=round,fill=fillColor,] (219.18,204.53) circle (  1.12);

\draw[color=drawColor,line cap=round,line join=round,fill=fillColor,] (220.86,202.95) circle (  1.12);

\draw[color=drawColor,line cap=round,line join=round,fill=fillColor,] (222.53,200.23) circle (  1.12);

\draw[color=drawColor,line cap=round,line join=round,fill=fillColor,] (224.20,199.63) circle (  1.12);

\draw[color=drawColor,line cap=round,line join=round,fill=fillColor,] (225.88,197.78) circle (  1.12);

\draw[color=drawColor,line cap=round,line join=round,fill=fillColor,] (227.55,197.71) circle (  1.12);

\draw[color=drawColor,line cap=round,line join=round,fill=fillColor,] (229.22,196.75) circle (  1.12);

\draw[color=drawColor,line cap=round,line join=round,fill=fillColor,] (230.90,196.11) circle (  1.12);

\draw[color=drawColor,line cap=round,line join=round,fill=fillColor,] (232.57,195.42) circle (  1.12);

\draw[color=drawColor,line cap=round,line join=round,fill=fillColor,] (234.24,192.34) circle (  1.12);

\draw[color=drawColor,line cap=round,line join=round,fill=fillColor,] (235.92,191.07) circle (  1.12);

\draw[color=drawColor,line cap=round,line join=round,fill=fillColor,] (237.59,190.23) circle (  1.12);

\draw[color=drawColor,line cap=round,line join=round,fill=fillColor,] (239.26,189.97) circle (  1.12);

\draw[color=drawColor,line cap=round,line join=round,fill=fillColor,] (240.94,186.42) circle (  1.12);

\draw[color=drawColor,line cap=round,line join=round,fill=fillColor,] (242.61,185.97) circle (  1.12);

\draw[color=drawColor,line cap=round,line join=round,fill=fillColor,] (244.28,183.99) circle (  1.12);

\draw[color=drawColor,line cap=round,line join=round,fill=fillColor,] (245.96,183.64) circle (  1.12);

\draw[color=drawColor,line cap=round,line join=round,fill=fillColor,] (247.63,183.25) circle (  1.12);

\draw[color=drawColor,line cap=round,line join=round,fill=fillColor,] (249.30,181.53) circle (  1.12);

\draw[color=drawColor,line cap=round,line join=round,fill=fillColor,] (250.97,181.36) circle (  1.12);

\draw[color=drawColor,line cap=round,line join=round,fill=fillColor,] (252.65,176.77) circle (  1.12);

\draw[color=drawColor,line cap=round,line join=round,fill=fillColor,] (254.32,174.94) circle (  1.12);

\draw[color=drawColor,line cap=round,line join=round,fill=fillColor,] (255.99,172.60) circle (  1.12);

\draw[color=drawColor,line cap=round,line join=round,fill=fillColor,] (257.67,172.25) circle (  1.12);

\draw[color=drawColor,line cap=round,line join=round,fill=fillColor,] (259.34,169.63) circle (  1.12);

\draw[color=drawColor,line cap=round,line join=round,fill=fillColor,] (261.01,167.74) circle (  1.12);

\draw[color=drawColor,line cap=round,line join=round,fill=fillColor,] (262.69,167.34) circle (  1.12);

\draw[color=drawColor,line cap=round,line join=round,fill=fillColor,] (264.36,166.55) circle (  1.12);

\draw[color=drawColor,line cap=round,line join=round,fill=fillColor,] (266.03,166.54) circle (  1.12);

\draw[color=drawColor,line cap=round,line join=round,fill=fillColor,] (267.71,166.52) circle (  1.12);

\draw[color=drawColor,line cap=round,line join=round,fill=fillColor,] (269.38,165.99) circle (  1.12);

\draw[color=drawColor,line cap=round,line join=round,fill=fillColor,] (271.05,164.28) circle (  1.12);

\draw[color=drawColor,line cap=round,line join=round,fill=fillColor,] (272.73,159.49) circle (  1.12);

\draw[color=drawColor,line cap=round,line join=round,fill=fillColor,] (274.40,153.84) circle (  1.12);

\draw[color=drawColor,line cap=round,line join=round,fill=fillColor,] (276.07,152.70) circle (  1.12);

\draw[color=drawColor,line cap=round,line join=round,fill=fillColor,] (277.75,151.36) circle (  1.12);

\draw[color=drawColor,line cap=round,line join=round,fill=fillColor,] (279.42,149.50) circle (  1.12);

\draw[color=drawColor,line cap=round,line join=round,fill=fillColor,] (281.09,149.24) circle (  1.12);

\draw[color=drawColor,line cap=round,line join=round,fill=fillColor,] (282.76,146.88) circle (  1.12);

\draw[color=drawColor,line cap=round,line join=round,fill=fillColor,] (284.44,145.96) circle (  1.12);

\draw[color=drawColor,line cap=round,line join=round,fill=fillColor,] (286.11,145.63) circle (  1.12);

\draw[color=drawColor,line cap=round,line join=round,fill=fillColor,] (287.78,143.50) circle (  1.12);

\draw[color=drawColor,line cap=round,line join=round,fill=fillColor,] (289.46,141.68) circle (  1.12);

\draw[color=drawColor,line cap=round,line join=round,fill=fillColor,] (291.13,140.32) circle (  1.12);

\draw[color=drawColor,line cap=round,line join=round,fill=fillColor,] (292.80,138.80) circle (  1.12);

\draw[color=drawColor,line cap=round,line join=round,fill=fillColor,] (294.48,137.89) circle (  1.12);

\draw[color=drawColor,line cap=round,line join=round,fill=fillColor,] (296.15,136.17) circle (  1.12);

\draw[color=drawColor,line cap=round,line join=round,fill=fillColor,] (297.82,135.62) circle (  1.12);

\draw[color=drawColor,line cap=round,line join=round,fill=fillColor,] (299.50,131.72) circle (  1.12);

\draw[color=drawColor,line cap=round,line join=round,fill=fillColor,] (301.17,130.28) circle (  1.12);

\draw[color=drawColor,line cap=round,line join=round,fill=fillColor,] (302.84,127.87) circle (  1.12);

\draw[color=drawColor,line cap=round,line join=round,fill=fillColor,] (304.52,126.20) circle (  1.12);

\draw[color=drawColor,line cap=round,line join=round,fill=fillColor,] (306.19,104.05) circle (  1.12);
\end{scope}
\begin{scope}
\path[clip] (  0.00,  0.00) rectangle (361.35,361.35);
\definecolor[named]{drawColor}{rgb}{0.00,0.00,0.00}

\node[color=drawColor,anchor=base,inner sep=0pt, outer sep=0pt, scale=  3.00] at (216.67, 21.60) {Collection};

\node[rotate= 90.00,color=drawColor,anchor=base,inner sep=0pt, outer sep=0pt, scale=  3.00] at ( 57.60,204.67) {Estimated size};
\end{scope}
\begin{scope}
\path[clip] (  0.00,  0.00) rectangle (361.35,361.35);
\definecolor[named]{drawColor}{rgb}{0.00,0.00,0.00}

\draw[color=drawColor,line cap=round,line join=round,fill opacity=0.00,] (120.00, 96.00) -- (120.00,313.35);

\draw[color=drawColor,line cap=round,line join=round,fill opacity=0.00,] (120.00,151.41) -- (114.00,151.41);

\draw[color=drawColor,line cap=round,line join=round,fill opacity=0.00,] (120.00,225.37) -- (114.00,225.37);

\draw[color=drawColor,line cap=round,line join=round,fill opacity=0.00,] (120.00,299.34) -- (114.00,299.34);

\node[rotate= 90.00,color=drawColor,anchor=base,inner sep=0pt, outer sep=0pt, scale=  2.50] at ( 96.00,151.41) {1e+03};

\node[rotate= 90.00,color=drawColor,anchor=base,inner sep=0pt, outer sep=0pt, scale=  2.50] at ( 96.00,299.34) {1e+09};

\draw[color=drawColor,line cap=round,line join=round,fill opacity=0.00,] (127.16, 96.00) -- (306.19, 96.00);

\draw[color=drawColor,line cap=round,line join=round,fill opacity=0.00,] (127.16, 96.00) -- (127.16, 90.00);

\draw[color=drawColor,line cap=round,line join=round,fill opacity=0.00,] (306.19, 96.00) -- (306.19, 90.00);

\node[color=drawColor,anchor=base,inner sep=0pt, outer sep=0pt, scale=  2.50] at (127.16, 60.00) {1};

\node[color=drawColor,anchor=base,inner sep=0pt, outer sep=0pt, scale=  2.50] at (306.19, 60.00) {108};
\end{scope}
\end{tikzpicture}

}
}

\hspace{-2mm}
\subfigure[QueryPools - Zipf]{
\scalebox{0.30}{

\begin{tikzpicture}[x=1pt,y=1pt]
\definecolor[named]{drawColor}{rgb}{0.00,0.00,0.00}
\definecolor[named]{fillColor}{rgb}{1.00,1.00,1.00}
\fill[color=fillColor,] (0,0) rectangle (361.35,361.35);
\begin{scope}
\path[clip] (108.00, 96.00) rectangle (313.35,313.35);
\definecolor[named]{drawColor}{rgb}{0.00,0.00,0.00}
\definecolor[named]{fillColor}{rgb}{0.00,0.00,0.00}

\draw[color=drawColor,line cap=round,line join=round,fill=fillColor,] (115.61,305.30) circle (  1.12);

\draw[color=drawColor,line cap=round,line join=round,fill=fillColor,] (117.38,304.57) circle (  1.12);

\draw[color=drawColor,line cap=round,line join=round,fill=fillColor,] (119.16,303.75) circle (  1.12);

\draw[color=drawColor,line cap=round,line join=round,fill=fillColor,] (120.94,302.77) circle (  1.12);

\draw[color=drawColor,line cap=round,line join=round,fill=fillColor,] (122.71,301.66) circle (  1.12);

\draw[color=drawColor,line cap=round,line join=round,fill=fillColor,] (124.49,299.73) circle (  1.12);

\draw[color=drawColor,line cap=round,line join=round,fill=fillColor,] (126.27,297.92) circle (  1.12);

\draw[color=drawColor,line cap=round,line join=round,fill=fillColor,] (128.04,294.78) circle (  1.12);

\draw[color=drawColor,line cap=round,line join=round,fill=fillColor,] (129.82,292.86) circle (  1.12);

\draw[color=drawColor,line cap=round,line join=round,fill=fillColor,] (131.60,289.70) circle (  1.12);

\draw[color=drawColor,line cap=round,line join=round,fill=fillColor,] (133.38,287.84) circle (  1.12);

\draw[color=drawColor,line cap=round,line join=round,fill=fillColor,] (135.15,287.70) circle (  1.12);

\draw[color=drawColor,line cap=round,line join=round,fill=fillColor,] (136.93,287.68) circle (  1.12);

\draw[color=drawColor,line cap=round,line join=round,fill=fillColor,] (138.71,287.58) circle (  1.12);

\draw[color=drawColor,line cap=round,line join=round,fill=fillColor,] (140.48,287.32) circle (  1.12);

\draw[color=drawColor,line cap=round,line join=round,fill=fillColor,] (142.26,286.48) circle (  1.12);

\draw[color=drawColor,line cap=round,line join=round,fill=fillColor,] (144.04,285.55) circle (  1.12);

\draw[color=drawColor,line cap=round,line join=round,fill=fillColor,] (145.81,284.67) circle (  1.12);

\draw[color=drawColor,line cap=round,line join=round,fill=fillColor,] (147.59,283.84) circle (  1.12);

\draw[color=drawColor,line cap=round,line join=round,fill=fillColor,] (149.37,283.66) circle (  1.12);

\draw[color=drawColor,line cap=round,line join=round,fill=fillColor,] (151.15,283.27) circle (  1.12);

\draw[color=drawColor,line cap=round,line join=round,fill=fillColor,] (152.92,282.44) circle (  1.12);

\draw[color=drawColor,line cap=round,line join=round,fill=fillColor,] (154.70,280.71) circle (  1.12);

\draw[color=drawColor,line cap=round,line join=round,fill=fillColor,] (156.48,278.58) circle (  1.12);

\draw[color=drawColor,line cap=round,line join=round,fill=fillColor,] (158.25,278.13) circle (  1.12);

\draw[color=drawColor,line cap=round,line join=round,fill=fillColor,] (160.03,277.99) circle (  1.12);

\draw[color=drawColor,line cap=round,line join=round,fill=fillColor,] (161.81,277.52) circle (  1.12);

\draw[color=drawColor,line cap=round,line join=round,fill=fillColor,] (163.58,276.81) circle (  1.12);

\draw[color=drawColor,line cap=round,line join=round,fill=fillColor,] (165.36,276.52) circle (  1.12);

\draw[color=drawColor,line cap=round,line join=round,fill=fillColor,] (167.14,276.42) circle (  1.12);

\draw[color=drawColor,line cap=round,line join=round,fill=fillColor,] (168.92,275.44) circle (  1.12);

\draw[color=drawColor,line cap=round,line join=round,fill=fillColor,] (170.69,274.90) circle (  1.12);

\draw[color=drawColor,line cap=round,line join=round,fill=fillColor,] (172.47,274.77) circle (  1.12);

\draw[color=drawColor,line cap=round,line join=round,fill=fillColor,] (174.25,274.66) circle (  1.12);

\draw[color=drawColor,line cap=round,line join=round,fill=fillColor,] (176.02,273.95) circle (  1.12);

\draw[color=drawColor,line cap=round,line join=round,fill=fillColor,] (177.80,273.61) circle (  1.12);

\draw[color=drawColor,line cap=round,line join=round,fill=fillColor,] (179.58,272.50) circle (  1.12);

\draw[color=drawColor,line cap=round,line join=round,fill=fillColor,] (181.35,270.28) circle (  1.12);

\draw[color=drawColor,line cap=round,line join=round,fill=fillColor,] (183.13,266.46) circle (  1.12);

\draw[color=drawColor,line cap=round,line join=round,fill=fillColor,] (184.91,266.01) circle (  1.12);

\draw[color=drawColor,line cap=round,line join=round,fill=fillColor,] (186.69,265.57) circle (  1.12);

\draw[color=drawColor,line cap=round,line join=round,fill=fillColor,] (188.46,265.35) circle (  1.12);

\draw[color=drawColor,line cap=round,line join=round,fill=fillColor,] (190.24,265.23) circle (  1.12);

\draw[color=drawColor,line cap=round,line join=round,fill=fillColor,] (192.02,260.08) circle (  1.12);

\draw[color=drawColor,line cap=round,line join=round,fill=fillColor,] (193.79,258.73) circle (  1.12);

\draw[color=drawColor,line cap=round,line join=round,fill=fillColor,] (195.57,258.52) circle (  1.12);

\draw[color=drawColor,line cap=round,line join=round,fill=fillColor,] (197.35,257.55) circle (  1.12);

\draw[color=drawColor,line cap=round,line join=round,fill=fillColor,] (199.12,253.31) circle (  1.12);

\draw[color=drawColor,line cap=round,line join=round,fill=fillColor,] (200.90,252.34) circle (  1.12);

\draw[color=drawColor,line cap=round,line join=round,fill=fillColor,] (202.68,252.25) circle (  1.12);

\draw[color=drawColor,line cap=round,line join=round,fill=fillColor,] (204.46,250.01) circle (  1.12);

\draw[color=drawColor,line cap=round,line join=round,fill=fillColor,] (206.23,249.31) circle (  1.12);

\draw[color=drawColor,line cap=round,line join=round,fill=fillColor,] (208.01,248.74) circle (  1.12);

\draw[color=drawColor,line cap=round,line join=round,fill=fillColor,] (209.79,248.49) circle (  1.12);

\draw[color=drawColor,line cap=round,line join=round,fill=fillColor,] (211.56,247.15) circle (  1.12);

\draw[color=drawColor,line cap=round,line join=round,fill=fillColor,] (213.34,247.08) circle (  1.12);

\draw[color=drawColor,line cap=round,line join=round,fill=fillColor,] (215.12,241.41) circle (  1.12);

\draw[color=drawColor,line cap=round,line join=round,fill=fillColor,] (216.89,240.84) circle (  1.12);

\draw[color=drawColor,line cap=round,line join=round,fill=fillColor,] (218.67,240.15) circle (  1.12);

\draw[color=drawColor,line cap=round,line join=round,fill=fillColor,] (220.45,238.63) circle (  1.12);

\draw[color=drawColor,line cap=round,line join=round,fill=fillColor,] (222.23,237.37) circle (  1.12);

\draw[color=drawColor,line cap=round,line join=round,fill=fillColor,] (224.00,236.30) circle (  1.12);

\draw[color=drawColor,line cap=round,line join=round,fill=fillColor,] (225.78,235.67) circle (  1.12);

\draw[color=drawColor,line cap=round,line join=round,fill=fillColor,] (227.56,232.57) circle (  1.12);

\draw[color=drawColor,line cap=round,line join=round,fill=fillColor,] (229.33,231.46) circle (  1.12);

\draw[color=drawColor,line cap=round,line join=round,fill=fillColor,] (231.11,231.11) circle (  1.12);

\draw[color=drawColor,line cap=round,line join=round,fill=fillColor,] (232.89,230.84) circle (  1.12);

\draw[color=drawColor,line cap=round,line join=round,fill=fillColor,] (234.66,230.42) circle (  1.12);

\draw[color=drawColor,line cap=round,line join=round,fill=fillColor,] (236.44,228.41) circle (  1.12);

\draw[color=drawColor,line cap=round,line join=round,fill=fillColor,] (238.22,227.13) circle (  1.12);

\draw[color=drawColor,line cap=round,line join=round,fill=fillColor,] (240.00,225.64) circle (  1.12);

\draw[color=drawColor,line cap=round,line join=round,fill=fillColor,] (241.77,223.60) circle (  1.12);

\draw[color=drawColor,line cap=round,line join=round,fill=fillColor,] (243.55,222.95) circle (  1.12);

\draw[color=drawColor,line cap=round,line join=round,fill=fillColor,] (245.33,222.21) circle (  1.12);

\draw[color=drawColor,line cap=round,line join=round,fill=fillColor,] (247.10,218.68) circle (  1.12);

\draw[color=drawColor,line cap=round,line join=round,fill=fillColor,] (248.88,216.87) circle (  1.12);

\draw[color=drawColor,line cap=round,line join=round,fill=fillColor,] (250.66,216.17) circle (  1.12);

\draw[color=drawColor,line cap=round,line join=round,fill=fillColor,] (252.43,213.99) circle (  1.12);

\draw[color=drawColor,line cap=round,line join=round,fill=fillColor,] (254.21,212.72) circle (  1.12);

\draw[color=drawColor,line cap=round,line join=round,fill=fillColor,] (255.99,212.67) circle (  1.12);

\draw[color=drawColor,line cap=round,line join=round,fill=fillColor,] (257.77,212.02) circle (  1.12);

\draw[color=drawColor,line cap=round,line join=round,fill=fillColor,] (259.54,211.33) circle (  1.12);

\draw[color=drawColor,line cap=round,line join=round,fill=fillColor,] (261.32,209.45) circle (  1.12);

\draw[color=drawColor,line cap=round,line join=round,fill=fillColor,] (263.10,207.72) circle (  1.12);

\draw[color=drawColor,line cap=round,line join=round,fill=fillColor,] (264.87,207.57) circle (  1.12);

\draw[color=drawColor,line cap=round,line join=round,fill=fillColor,] (266.65,205.27) circle (  1.12);

\draw[color=drawColor,line cap=round,line join=round,fill=fillColor,] (268.43,203.18) circle (  1.12);

\draw[color=drawColor,line cap=round,line join=round,fill=fillColor,] (270.20,202.74) circle (  1.12);

\draw[color=drawColor,line cap=round,line join=round,fill=fillColor,] (271.98,201.30) circle (  1.12);

\draw[color=drawColor,line cap=round,line join=round,fill=fillColor,] (273.76,199.43) circle (  1.12);

\draw[color=drawColor,line cap=round,line join=round,fill=fillColor,] (275.54,199.37) circle (  1.12);

\draw[color=drawColor,line cap=round,line join=round,fill=fillColor,] (277.31,197.63) circle (  1.12);

\draw[color=drawColor,line cap=round,line join=round,fill=fillColor,] (279.09,196.14) circle (  1.12);

\draw[color=drawColor,line cap=round,line join=round,fill=fillColor,] (280.87,193.75) circle (  1.12);

\draw[color=drawColor,line cap=round,line join=round,fill=fillColor,] (282.64,191.71) circle (  1.12);

\draw[color=drawColor,line cap=round,line join=round,fill=fillColor,] (284.42,189.30) circle (  1.12);

\draw[color=drawColor,line cap=round,line join=round,fill=fillColor,] (286.20,186.07) circle (  1.12);

\draw[color=drawColor,line cap=round,line join=round,fill=fillColor,] (287.97,181.67) circle (  1.12);

\draw[color=drawColor,line cap=round,line join=round,fill=fillColor,] (289.75,179.11) circle (  1.12);

\draw[color=drawColor,line cap=round,line join=round,fill=fillColor,] (291.53,176.19) circle (  1.12);

\draw[color=drawColor,line cap=round,line join=round,fill=fillColor,] (293.31,172.77) circle (  1.12);

\draw[color=drawColor,line cap=round,line join=round,fill=fillColor,] (295.08,166.63) circle (  1.12);

\draw[color=drawColor,line cap=round,line join=round,fill=fillColor,] (296.86,163.34) circle (  1.12);

\draw[color=drawColor,line cap=round,line join=round,fill=fillColor,] (298.64,159.10) circle (  1.12);

\draw[color=drawColor,line cap=round,line join=round,fill=fillColor,] (300.41,155.12) circle (  1.12);

\draw[color=drawColor,line cap=round,line join=round,fill=fillColor,] (302.19,154.43) circle (  1.12);

\draw[color=drawColor,line cap=round,line join=round,fill=fillColor,] (303.97,138.55) circle (  1.12);

\draw[color=drawColor,line cap=round,line join=round,fill=fillColor,] (305.74,104.05) circle (  1.12);
\end{scope}
\begin{scope}
\path[clip] (  0.00,  0.00) rectangle (361.35,361.35);
\definecolor[named]{drawColor}{rgb}{0.00,0.00,0.00}

\node[color=drawColor,anchor=base,inner sep=0pt, outer sep=0pt, scale=  3.00] at (210.67, 21.60) {Collection%
};

\node[rotate= 90.00,color=drawColor,anchor=base,inner sep=0pt, outer sep=0pt, scale=  3.00] at ( 45.60,204.67) {Estimated size%
};
\end{scope}
\begin{scope}
\path[clip] (  0.00,  0.00) rectangle (361.35,361.35);
\definecolor[named]{drawColor}{rgb}{0.00,0.00,0.00}

\draw[color=drawColor,line cap=round,line join=round,fill opacity=0.00,] (108.00, 95.39) -- (108.00,306.69);

\draw[color=drawColor,line cap=round,line join=round,fill opacity=0.00,] (108.00,147.66) -- (102.00,147.66);

\draw[color=drawColor,line cap=round,line join=round,fill opacity=0.00,] (108.00,227.18) -- (102.00,227.18);

\draw[color=drawColor,line cap=round,line join=round,fill opacity=0.00,] (108.00,306.69) -- (102.00,306.69);

\node[rotate= 90.00,color=drawColor,anchor=base,inner sep=0pt, outer sep=0pt, scale=  2.50] at ( 84.00,147.66) {1e+02%
};

\node[rotate= 90.00,color=drawColor,anchor=base,inner sep=0pt, outer sep=0pt, scale=  2.50] at ( 84.00,306.69) {1e+06%
};

\draw[color=drawColor,line cap=round,line join=round,fill opacity=0.00,] (115.61, 96.00) -- (305.74, 96.00);

\draw[color=drawColor,line cap=round,line join=round,fill opacity=0.00,] (115.61, 96.00) -- (115.61, 90.00);

\draw[color=drawColor,line cap=round,line join=round,fill opacity=0.00,] (305.74, 96.00) -- (305.74, 90.00);

\node[color=drawColor,anchor=base,inner sep=0pt, outer sep=0pt, scale=  2.50] at (115.61, 60.00) {1%
};

\node[color=drawColor,anchor=base,inner sep=0pt, outer sep=0pt, scale=  2.50] at (305.74, 60.00) {108%
};
\end{scope}
\end{tikzpicture}

}
}
}

\caption{Size distribution}
\label{size distribution}
\end{center}
\end{figure*}

\subsection{Methods}
\noindent \textbf{ClueWeb I} 
Our first method is built on the intuition to scale document frequencies using a reference corpus. 
We will use the ClueWeb09 dataset that represents a crawl of the web. 
For each query $q \in Q$  obtained using query based sampling,
we calculate the size by dividing the number of results by its ClueWeb document frequency $df_{q,Clueweb}$ and scaling by the total
ClueWeb size $|C_{Clueweb}|$.  We then take the average of all queries to estimate the size $\hat{N}_i$ of collection $C_i$.

\[\hat{N}_i = \frac{1}{|Q|} \sum_{q \in Q} \frac{numResults_{q,i}}{df_{q,Clueweb}} \times |C_{Clueweb}|\]

The method is limited by the maximum number of results that is collected for the sample queries.
Here, at most 10 results were collected per query, giving a maximum estimate of 10 times the ClueWeb size.
However, a big advantage is that this maximum estimate is independent of the number of queries.
The choice of reference corpus is important though, and choosing the ClueWeb09 dataset will 
estimate large sizes for collections that do well on most terms (have a broad coverage, just as the web).
Very specific collections that only return results for a couple of queries, but are in fact very large,
will have a low estimate using this method. 
This method is not robust against resources that always return the maximum number of results (regardless whether the results match the query or not),
since it will then incorrectly give a very high estimate. \\

\noindent \textbf{ClueWeb II} 
The method using ClueWeb09 gives low estimates for very frequent terms such as \emph{the} due to the limited number of results for each
sample query (maximum of 10). 
Therefore, we also explore a variant, that discards sample queries for which a resource returned 10 results
(the maximum number) while the ClueWeb09 corpus has a higher document frequency for that term. 
\\

\noindent \textbf{QueryPools} Our second method is based on query pools and a modification of the method used by Broder et al. \cite{Broder:2006:ECS:1183614.1183699}.
 We divide the queries randomly into two sets to create two uncorrelated sets of queries ($A$ and $B$), and use the document overlap $|D_{AB}|$ to estimate the size 
 using $\frac{|D_A||D_B|}{|D_{AB}|}$ \cite{Liu:2002:DRS:584792.584909}.
This method has two limitations. First, if there is no document overlap, the result is not well defined.
In our estimates, we then take $|D_{AB}|$ to be 1. Second,
the actual estimate is limited by the number of samples.
When taking the minimum value of $|D_{AB}|$ to be 1, the maximum value is $|D_A||D_B|$.
For example, to get an estimate of 600 mln, $|D_A|$ and $|D_B|$ should be 
24.60 mln (assuming no overlap), an infeasible amount of samples.
 \\

 \subsection{Analysis}
  The \emph{ClueWeb} method uses an external resource to estimate the size, while the \emph{QueryPools} method
uses the overlapping documents. The size distributions of the methods are plotted in Figure \ref{size distribution}
and examples of estimated sizes are presented in Table \ref{example sizes}. 
We find a very skewed size distribution using all methods. Because these search engines are real search engines on the
web, their size is unknown. However, the results do suggest that the sizes are extremely skewed, much more
than in previous test collections for federated search.
Note that the exact size estimates differ by orders of magnitude when comparing both methods, with the ClueWeb method giving much higher estimates.
The reason of this was outlined in the previous section.

\begin{table*}[ht]
\centering
\begin{tabular}{llllll}
\hline  \multicolumn{2}{c}{\textbf{ClueWeb I  - Zipf}} & \multicolumn{2}{c}{\textbf{ClueWeb II - Zipf}} &  \multicolumn{2}{c}{\textbf{QueryPools - Zipf}} \\
 \hline Google	 & 562 mln &  Baidu & 1,745 mln & Google & 923k\\
 Mamma.com & 526 mln  & Babylon & 889 mln & Baidu & 884k\\
 Yahoo & 521 mln & Bing  & 715 mln  &  Bing & 843k\\
 Baidu & 495 mln & Google & 621 mln & Ask & 797k\\
 IMDb & 485 mln & Google Blogs & 503 mln & Google Books & 747k\\
\hline
\end{tabular}
\caption{Largest resources}
\label{example sizes}
\end{table*}

\section{Resource Selection}
We first outline the experimental setup. Then, we  demonstrate the potential of a federated search system using the resource selection task. Next, we test the performance of  well known resource selection methods on this dataset and discuss the results.

\subsection{Experimental setup}
\noindent \textbf{Task} 
For a given query $q$, the goal of resource selection is to select the top $k$ best suited resources to search.\\

\noindent \textbf{Evaluation} 
We evaluate our methods using two metrics. \\

\noindent \emph{recall@k} (also used by
\cite{Si:2003:RDD:860435.860490,Si:2004:UUM:1031171.1031180,Shokouhi:2007:CCS:1763653.1763674})
is the proportion of relevant documents in the top \emph{k} collections
of the resulting ranking $\Omega$ compared to the top \emph{k} collections of the optimal ranking \emph{O}
(in this paper $k$ is usually 5):

\[Recall = R_k = \frac{\sum_{i=1}^k \Omega_i}{\sum_{i=1}^k O_i}\]

\noindent \emph{precision@(10$\times k$)} Note that with the recall metric relevant documents returned by multiple collections are counted as relevant multiple times. To account for overlapping documents, we report the precision@(10$\times k$), by assuming that each resource returns 10 results and calculating the precision after merging the results.
By using the above cutoff, we overcome the problem of resource merging, we simply aggregate all results
for the top $k$ collections. Duplicate documents are only counted once (all others are considered irrelevant).  
This is similar to the evaluation method used by Shokouhi and Zobel \cite{Shokouhi:2007:FTR:1277741.1277827}, however
they did use a resource merging method.\\

\noindent \textbf{URL normalization}  Whether documents returned by different resources are in fact the same, is decided based on their URLs.
Therefore, URLs were first normalized to a standard form, by deleting added strings such as search engine specific additions
and query terms.\\

\noindent \textbf{Binary relevance} The page judgements were done on a six point scale. However,
in this paper we will use binary relevance to evaluate the resource selection task. 
We have listed the proportions of page relevance categories
in the annotations (without normalizing the URLs) in Table ~\ref{page-judgements}. 
Unless specified otherwise, we will consider a page to be relevant when on average it was ranked higher than \emph{Rel}.
We also sometimes
use the very strict criterion higher than \emph{Hrel}, such that less than 3 percent of the results is considered relevant.

 \begin{table}[ht]
\centering
\begin{tabular}{rc}
\hline
\textbf{Page relevance} & \textbf{Percentage} \\
\hline  
Junk/Non & 82.841 \\
Rel & 8.756\\
HRel &  5.533\\
Key & 2.456\\
 Nav & 0.413 \\
\hline
\end{tabular}
\caption{Page judgements}
\label{page-judgements}
\end{table}

\subsection{Web search without the general web search engines?}
In this section our goal is to analyze what performance can be obtained without the large web search engines
in the context of resource selection
\footnote{For our comparison between WSE's and non-WSE's, we only consider the potential number of relevant results; clearly other parameters play a role as well, e.g., the combined non-WSE's are most likely not capable of processing the huge workload of the WSE's.}.
We divide the resources into two sets: 

\begin{enumerate}
\item  \emph{The general web search engines} (WSE's), containing 10 resources: Google, Bing, Yahoo, AOL, Ask, Mamma.com, Gigablast, Blekko, Baidu, Babylon.
\item \emph{The remaining search engines} (indicated as non-WSE's), containing 98 resources. Many of them
are very small, however some of these non-WSE's can still be very large, such as search engines of large sites
such as YouTube, Flickr and Amazon, or specialized search engines like Google Blogs.
\end{enumerate}

For each query, an oracle ranking for both resource sets was created based
on the number of relevant documents. The recall is calculated by dividing by the 
optimal ranking with \emph{all}
 search engines included. Results can be found in Table \ref{oracle results}.
Note that when evaluating at $k=5$ we also select the most optimal resources for
the WSE's (according to the number of relevant documents).

 \begin{table*}[ht]
\centering
\begin{tabular}{ccccc}
\hline  &\multicolumn{2}{c}{ \textbf{Only WSE}}  	 &\multicolumn{2}{c}{ \textbf{Non-WSE}} 	\\
 & \textbf{Precision} & \textbf{Recall} & \textbf{Precision} & \textbf{Recall} \\
\hline k = 5 & \specialcell{All: 0.248 \\Ambiguous:  0.198 \\ Faceted: 0.307} & \specialcell{All: 0.828\\Ambiguous:  0.769\\ Faceted: 0.897 }  &	\specialcell{All:  0.343 \\Ambiguous:  0.304 \\ Faceted:  0.388}  & \specialcell{All:  0.619\\Ambiguous: 0.671 \\ Faceted: 0.558 }	\\ 
\hline k = 10 &  \specialcell{All:  0.153\\Ambiguous: 0.113 \\ Faceted: 0.201} & \specialcell{All: 0.742\\Ambiguous:  0.674\\ Faceted: 0.822 } &	 \specialcell{All: 0.231 \\Ambiguous: 0.204 \\ Faceted: 0.263 }& \specialcell{All: 0.514 \\Ambiguous: 0.568  \\ Faceted: 0.449}\\ 
\hline
\end{tabular}
\caption{Oracle experiment}
\label{oracle results}
\end{table*}

We find that for the recall metric the run using only general web search engines (WSE's)
achieves higher performance, indicating that this system is able to achieve a higher total number
of relevant documents. However, when we use the precision metric, which does
take the unique number of results into account, the run without general web search engines performs better.
We find that the WSE's return a lot of duplicate relevant documents.

 Figure \ref{Comparison oracles} shows a plot with the difference for precision per topic between the two sets.
 A positive value indicates that the oracle ranking using only WSE's performed better.
\begin{figure}[htb]
\center
\scalebox{0.4}{

\begin{tikzpicture}[x=1pt,y=1pt]
\definecolor[named]{drawColor}{rgb}{0.00,0.00,0.00}
\definecolor[named]{fillColor}{rgb}{1.00,1.00,1.00}
\fill[color=fillColor,fill opacity=0.00,] (0,0) rectangle (361.35,361.35);
\begin{scope}
\path[clip] (  0.00,  0.00) rectangle (361.35,361.35);
\definecolor[named]{drawColor}{rgb}{0.00,0.00,0.00}
\definecolor[named]{fillColor}{rgb}{0.30,0.30,0.30}

\draw[color=drawColor,line cap=round,line join=round,fill=fillColor,] ( 59.83,237.54) rectangle ( 64.27,312.15);

\draw[color=drawColor,line cap=round,line join=round,fill=fillColor,] ( 65.16,237.54) rectangle ( 69.60,298.59);

\draw[color=drawColor,line cap=round,line join=round,fill=fillColor,] ( 70.49,237.54) rectangle ( 74.93,298.59);

\draw[color=drawColor,line cap=round,line join=round,fill=fillColor,] ( 75.82,237.54) rectangle ( 80.27,291.80);

\draw[color=drawColor,line cap=round,line join=round,fill=fillColor,] ( 81.15,237.54) rectangle ( 85.60,285.02);

\draw[color=drawColor,line cap=round,line join=round,fill=fillColor,] ( 86.49,237.54) rectangle ( 90.93,278.24);

\draw[color=drawColor,line cap=round,line join=round,fill=fillColor,] ( 91.82,237.54) rectangle ( 96.26,271.46);

\draw[color=drawColor,line cap=round,line join=round,fill=fillColor,] ( 97.15,237.54) rectangle (101.59,271.46);

\draw[color=drawColor,line cap=round,line join=round,fill=fillColor,] (102.48,237.54) rectangle (106.92,271.46);

\draw[color=drawColor,line cap=round,line join=round,fill=fillColor,] (107.81,237.54) rectangle (112.26,264.67);

\draw[color=drawColor,line cap=round,line join=round,fill=fillColor,] (113.14,237.54) rectangle (117.59,257.89);

\draw[color=drawColor,line cap=round,line join=round,fill=fillColor,] (118.48,237.54) rectangle (122.92,257.89);

\draw[color=drawColor,line cap=round,line join=round,fill=fillColor,] (123.81,237.54) rectangle (128.25,251.11);

\draw[color=drawColor,line cap=round,line join=round,fill=fillColor,] (129.14,237.54) rectangle (133.58,251.11);

\draw[color=drawColor,line cap=round,line join=round,fill=fillColor,] (134.47,237.54) rectangle (138.91,244.33);

\draw[color=drawColor,line cap=round,line join=round,fill=fillColor,] (139.80,237.54) rectangle (144.25,237.54);

\draw[color=drawColor,line cap=round,line join=round,fill=fillColor,] (145.13,237.54) rectangle (149.58,237.54);

\draw[color=drawColor,line cap=round,line join=round,fill=fillColor,] (150.47,237.54) rectangle (154.91,237.54);

\draw[color=drawColor,line cap=round,line join=round,fill=fillColor,] (155.80,237.54) rectangle (160.24,237.54);

\draw[color=drawColor,line cap=round,line join=round,fill=fillColor,] (161.13,237.54) rectangle (165.57,230.76);

\draw[color=drawColor,line cap=round,line join=round,fill=fillColor,] (166.46,237.54) rectangle (170.90,230.76);

\draw[color=drawColor,line cap=round,line join=round,fill=fillColor,] (171.79,237.54) rectangle (176.24,230.76);

\draw[color=drawColor,line cap=round,line join=round,fill=fillColor,] (177.12,237.54) rectangle (181.57,223.98);

\draw[color=drawColor,line cap=round,line join=round,fill=fillColor,] (182.46,237.54) rectangle (186.90,223.98);

\draw[color=drawColor,line cap=round,line join=round,fill=fillColor,] (187.79,237.54) rectangle (192.23,217.20);

\draw[color=drawColor,line cap=round,line join=round,fill=fillColor,] (193.12,237.54) rectangle (197.56,210.41);

\draw[color=drawColor,line cap=round,line join=round,fill=fillColor,] (198.45,237.54) rectangle (202.89,203.63);

\draw[color=drawColor,line cap=round,line join=round,fill=fillColor,] (203.78,237.54) rectangle (208.23,196.85);

\draw[color=drawColor,line cap=round,line join=round,fill=fillColor,] (209.11,237.54) rectangle (213.56,196.85);

\draw[color=drawColor,line cap=round,line join=round,fill=fillColor,] (214.45,237.54) rectangle (218.89,196.85);

\draw[color=drawColor,line cap=round,line join=round,fill=fillColor,] (219.78,237.54) rectangle (224.22,196.85);

\draw[color=drawColor,line cap=round,line join=round,fill=fillColor,] (225.11,237.54) rectangle (229.55,190.07);

\draw[color=drawColor,line cap=round,line join=round,fill=fillColor,] (230.44,237.54) rectangle (234.88,183.28);

\draw[color=drawColor,line cap=round,line join=round,fill=fillColor,] (235.77,237.54) rectangle (240.22,176.50);

\draw[color=drawColor,line cap=round,line join=round,fill=fillColor,] (241.10,237.54) rectangle (245.55,169.72);

\draw[color=drawColor,line cap=round,line join=round,fill=fillColor,] (246.44,237.54) rectangle (250.88,169.72);

\draw[color=drawColor,line cap=round,line join=round,fill=fillColor,] (251.77,237.54) rectangle (256.21,169.72);

\draw[color=drawColor,line cap=round,line join=round,fill=fillColor,] (257.10,237.54) rectangle (261.54,162.94);

\draw[color=drawColor,line cap=round,line join=round,fill=fillColor,] (262.43,237.54) rectangle (266.87,162.94);

\draw[color=drawColor,line cap=round,line join=round,fill=fillColor,] (267.76,237.54) rectangle (272.21,162.94);

\draw[color=drawColor,line cap=round,line join=round,fill=fillColor,] (273.09,237.54) rectangle (277.54,156.15);

\draw[color=drawColor,line cap=round,line join=round,fill=fillColor,] (278.43,237.54) rectangle (282.87,149.37);

\draw[color=drawColor,line cap=round,line join=round,fill=fillColor,] (283.76,237.54) rectangle (288.20,142.59);

\draw[color=drawColor,line cap=round,line join=round,fill=fillColor,] (289.09,237.54) rectangle (293.53,129.02);

\draw[color=drawColor,line cap=round,line join=round,fill=fillColor,] (294.42,237.54) rectangle (298.86,115.46);

\draw[color=drawColor,line cap=round,line join=round,fill=fillColor,] (299.75,237.54) rectangle (304.20,101.89);

\draw[color=drawColor,line cap=round,line join=round,fill=fillColor,] (305.08,237.54) rectangle (309.53, 95.11);

\draw[color=drawColor,line cap=round,line join=round,fill=fillColor,] (310.42,237.54) rectangle (314.86, 81.55);

\draw[color=drawColor,line cap=round,line join=round,fill=fillColor,] (315.75,237.54) rectangle (320.19, 74.76);

\draw[color=drawColor,line cap=round,line join=round,fill=fillColor,] (321.08,237.54) rectangle (325.52, 61.20);
\end{scope}
\begin{scope}
\path[clip] (  0.00,  0.00) rectangle (361.35,361.35);
\definecolor[named]{drawColor}{rgb}{0.00,0.00,0.00}

\draw[color=drawColor,line cap=round,line join=round,fill opacity=0.00,] ( 49.20, 67.98) -- ( 49.20,305.37);

\draw[color=drawColor,line cap=round,line join=round,fill opacity=0.00,] ( 49.20, 67.98) -- ( 43.20, 67.98);

\draw[color=drawColor,line cap=round,line join=round,fill opacity=0.00,] ( 49.20,101.89) -- ( 43.20,101.89);

\draw[color=drawColor,line cap=round,line join=round,fill opacity=0.00,] ( 49.20,135.81) -- ( 43.20,135.81);

\draw[color=drawColor,line cap=round,line join=round,fill opacity=0.00,] ( 49.20,169.72) -- ( 43.20,169.72);

\draw[color=drawColor,line cap=round,line join=round,fill opacity=0.00,] ( 49.20,203.63) -- ( 43.20,203.63);

\draw[color=drawColor,line cap=round,line join=round,fill opacity=0.00,] ( 49.20,237.54) -- ( 43.20,237.54);

\draw[color=drawColor,line cap=round,line join=round,fill opacity=0.00,] ( 49.20,271.46) -- ( 43.20,271.46);

\draw[color=drawColor,line cap=round,line join=round,fill opacity=0.00,] ( 49.20,305.37) -- ( 43.20,305.37);

\node[rotate= 90.00,color=drawColor,anchor=base,inner sep=0pt, outer sep=0pt, scale=  1.00] at ( 37.20, 67.98) {-0.5};

\node[rotate= 90.00,color=drawColor,anchor=base,inner sep=0pt, outer sep=0pt, scale=  1.00] at ( 37.20,101.89) {-0.4};

\node[rotate= 90.00,color=drawColor,anchor=base,inner sep=0pt, outer sep=0pt, scale=  1.00] at ( 37.20,135.81) {-0.3};

\node[rotate= 90.00,color=drawColor,anchor=base,inner sep=0pt, outer sep=0pt, scale=  1.00] at ( 37.20,169.72) {-0.2};

\node[rotate= 90.00,color=drawColor,anchor=base,inner sep=0pt, outer sep=0pt, scale=  1.00] at ( 37.20,203.63) {-0.1};

\node[rotate= 90.00,color=drawColor,anchor=base,inner sep=0pt, outer sep=0pt, scale=  1.00] at ( 37.20,237.54) {0.0};

\node[rotate= 90.00,color=drawColor,anchor=base,inner sep=0pt, outer sep=0pt, scale=  1.00] at ( 37.20,271.46) {0.1};

\node[rotate= 90.00,color=drawColor,anchor=base,inner sep=0pt, outer sep=0pt, scale=  1.00] at ( 37.20,305.37) {0.2};
\end{scope}
\begin{scope}
\path[clip] (  0.00,  0.00) rectangle (361.35,361.35);
\definecolor[named]{drawColor}{rgb}{0.00,0.00,0.00}

\node[color=drawColor,anchor=base,inner sep=0pt, outer sep=0pt, scale=  2.00] at (192.68, 13.20) {Precision of WSE vs. Non-WSE per topic};

\node[rotate= 90.00,color=drawColor,anchor=base,inner sep=0pt, outer sep=0pt, scale=  2.00] at ( 13.20,186.67) {Prec[WSE] - Prec[non-WSE]};
\end{scope}
\end{tikzpicture}

}
\caption{Comparison oracles}
\label{Comparison oracles}
\end{figure}
Table \ref{oracle results} also shows a breakdown for ambiguous versus faceted queries.
We find that the WSE's do especially well on faceted queries. 
The oracle system using the other search engines does well on ambiguous queries when looking at recall, because
in an oracle ranking the resources with the correct sense are ranked high. 

An example query for which the oracle ranking using non-WSE's achieved better performance is 
the highly ambiguous query
\emph{avp} (`Find information about events sponsored by AVP, the Association of
    Volleyball Professionals.').
The top 5 of resources for this query are Myspace, WordPress, ESPN, Metacafe	and Encyclopedia Britannica.
An example of a query for which the oracle ranking using only WSE's was much better is
\emph{president of the united states} (`Find information about the office of President of the United States.').
The top 5 resources for this query are Gigablast, Ask, Yahoo, Mamma.com and Bing, all
WSE's. 

Next, we use a more strict criterion of relevance. We only consider pages to be relevant when
they are on average judged higher than \emph{HRel}. The results can be found in Table ~\ref{oracle results hrel}.
Note that the average precision values are low,
because the queries are highly ambiguous and we use a very strict criterion. 
We now observe that
the general web search engines perform much better. For example, a system without general web search engines
might miss a lot of relevant pages for navigational queries. 
Also, most of the test queries are general requests for information, hence better suited for the WSE's, as becomes clear further on.

 \begin{table}[ht]
\centering
\begin{tabular}{ccccc}
\hline  &\multicolumn{2}{c}{ \textbf{Only WSE}}  	 &\multicolumn{2}{c}{ \textbf{Non-WSE}} 	\\
 & \textbf{Precision} & \textbf{Recall} & \textbf{Precision} & \textbf{Recall} \\
\hline k = 5 & 0.120 &  0.891 & 0.065  & 0.237	\\ 
\hline k = 10 &  0.070 & 0.844  & 0.033  & 	0.187\\ 
\hline
\end{tabular}
\caption{Oracle experiment - Better than HRel}
\label{oracle results hrel}
\end{table}

When we use a less strict criterion of relevance (Table ~\ref{oracle results rel}), we find the difference in precision between the oracles with 
WSE's and non-WSE's even bigger. Thus, the non-WSE's could be
an alternative to WSE's when taking a normal criterion on relevance. 
However, when the goal
is to deliver high relevant results for example to answer navigational queries, the WSE's are still better suited.

 \begin{table}[ht]
\centering
\begin{tabular}{ccccc}
\hline  &\multicolumn{2}{c}{ \textbf{Only WSE}}  	 &\multicolumn{2}{c}{ \textbf{Non-WSE}} 	\\
 & \textbf{Precision} & \textbf{Recall} & \textbf{Precision} & \textbf{Recall} \\
\hline k = 5 & 0.328  & 0.835   &  0.561  & 0.772 \\ 
\hline k = 10 & 0.217   & 0.735    &  0.437  & 0.684  \\ 
\hline
\end{tabular}
\caption{Oracle experiment - Rel or better}
\label{oracle results rel}
\end{table}

\subsection{Resource selection experiments}
As demonstrated before, an optimal resource selection system can be very powerful
and a possible alternative to the general web search engines. We now test
several well known resource selection methods for this task.

\subsubsection{Preprocessing}
Snippets were indexed using the text from the URL, title, and summary.
The pages were indexed using the URL and text extracted from the document.
Only text was extracted from HTML pages and PDF files. Some of the collections
returned media files such as images (jpg, png, gif, svg) and sounds (ogg),
for these files only the URL could be used for indexing.
Terrier \cite{ounis06terrier-osir} was used for the retrieval experiments.
No stopwords were removed, and words were stemmed using the Porter stemmer.

\subsubsection{Baselines}
\label{baselines section}
 We compare with the following baselines. 
\begin{itemize}
\item \emph{Popular} selects the top 5 most popular search engines in the US: Google, Bing, Yahoo, AOL and Ask \cite{comscore}.
\item \emph{Size Based 1} (SB1) ranks only those collections on size, that match at least one query term.
\item \emph{Size Based 2} (SB2) ranks all collections on size, resulting in the same ranking for each query.
\end{itemize}

\subsubsection{Big document models}

\noindent \textbf{CORI} is a big document model (\cite{Callan95searchingdistributed}, \cite{callan2000}) and
an adaptation of the INQUERY method. Collections
are represented by their terms and document frequencies and scored using:

\[p(w|C_i) = b + (1-b) \times T \times I\]

\noindent with

\[T =  \frac{df_{w,i}}{df_{w,i} + 50 + 150 \times \frac{cw_i}{avg_{cw}}}\]

\[I =  \frac{ \log(\frac{|C| + 0.5}{cf_q})}{\log(|C| + 1.0)}\]

\noindent where $|C|$ is the number of collections, 
$df_{w,i}$ is the number of documents in $C_i$ containing query term $w$,
$cw_i$ is the number of indexing terms in $C_i$,  
$avg_{cw}$ is the average number of indexing terms in each collection and 
$cf_q$ is the number of collections containing query term $w$.
The free parameter $b$ is usually set to 0.4. We set it using cross fold validation.
Collections are ranked by summing the beliefs of $P(w|C_i)$.\\

\noindent \textbf{LM} \cite{Si:2002:LMF:584792.584856}. A big document is created for 
each collection, and using language modeling with Jelinek-Mercer smoothing
the collections are ranked. $G$ is the global language model
by collapsing together all documents of all collections.
$\lambda$ is set using cross fold validation.

\[P(q|C_i) = \prod_{w \in q} \lambda P(w|C_i) + (1 - \lambda) P (w|G) \]

\subsubsection{Small document models}

\noindent \textbf{ReDDE} (Relevant Document Distribution Estimation method) \cite{Si:2003:RDD:860435.860490} 
estimates the number of relevant documents using
the following equation:

\[Rel(C_i,q) = \frac{|C_i|}{|S_i|} \times \sum_{d \in S_i} P(rel|d,q)\]

\noindent Where $|C_i|$ is the estimated size of collection $i$, and $|S_i|$ is the number of samples for
collection $i$.  A query is issued to the sample index, the rank of a document $d$ is then referred to as $sampleRank(d)$. ReDDE then estimates $centralRank(d)$, the rank 
 of the document in the complete centralized index.
A fixed probability $k$ is then assigned to the top ranked documents in the centralized index (in our experiments $k=1$).

\[P(rel|d,q) = k \mbox{ if }\mathit{centralRank}(d) < ratio * |C_{all}|\]

\noindent The $ratio$ is usually set within a range of 0.002 to 0.005. We set it using cross validation.
The centralized rank is estimated as follows:

\[centralRank(d) = \sum_{d', sampleRank(d') < sampleRank(d)} \frac{|C_{c_d'}|}{|S_{c_d'}|}\]

\noindent Where $|C_{c_d'}|$ is the size of the collection of document $d'$.\\

\noindent \textbf{GAVG} \cite{Seo:2008:BSS:1458082.1458222} scores a collection $C_i$
using the geometric average query likelihood from $C_i$'s top $k$ documents.
The query likelihood is calculated using Dirichlet smoothing. The smoothing parameter and $k$ are determined
using cross validation.
If a collection has ranked less than $k$ documents for the given query, the minimum query likelihood score is used
for the missing documents.

\[GAVG(C_i,q) = ( \prod_{d \in \mbox{top } k \mbox{ from } C_i} P(q|d))^{\frac{1}{k}}\]

\noindent \textbf{CRCS} (Central rank based collection selection) \cite{Shokouhi:2007:CCS:1763653.1763674}
scores a collection according to:

\[CRCS(C_i,q) = \frac{|C_i|}{|C^{max}| \times |S_i|} \times \sum_{d \in S_i} R(d))\]

\noindent with $|C_i|$ the estimated size of collection $C_i$, $|C^{max}|$ 
the size of the largest collection and $|S_i|$ the size of the sample.
$R(d)$ is nonzero when document $d$ is ranked in the top $k$ documents
of the sample index. In contrary to ReDDE, the relevance of a document $R(d)$ depends on its $sampleRank(d)$.
Note that the $sampleRank(d)$ is the same as used in ReDDE.
Two variations are proposed to calculate $R(d)$.

\[\mbox{Linear: } R(d) = k - sampleRank(d) \]

\[\mbox{Exponential: } R(d) = \alpha \times exp(-\beta \times sampleRank(d))\]

\noindent 
The parameters $k$, $\alpha$ and $\beta$ are set using cross fold validation.
Preliminary experiments showed no significant difference between the linear and exponential version, 
therefore we have only used the linear version in our final experiments.

\subsection{Results}
\noindent In the following paragraphs, we will give the results for the measures and experiments introduced in Section 4 and 5.\\

\noindent \textbf{Size} \\
\noindent The results are presented in Figure \ref{results}. We find that methods that explicitly take the size of the collection
into account achieve a much higher performance. Even a simple baseline, where all resources
that match at least one of the query terms are ranked on size (SB1), achieves a high performance compared to the other
methods. There are several possible explanations. The queries might be biased towards larger collections,
the resource selection methods might not be suitable for this setting and/or the samplings are not sufficient.
We analyze this further in the next section.\\

\begin{figure}[htb]
\center
\scalebox{0.7}{

\begin{tikzpicture}[x=1pt,y=1pt]
\definecolor[named]{drawColor}{rgb}{0.00,0.00,0.00}
\definecolor[named]{fillColor}{rgb}{1.00,1.00,1.00}
\fill[color=fillColor,fill opacity=0.00,] (0,0) rectangle (361.35,361.35);
\begin{scope}
\path[clip] (  0.00,  0.00) rectangle (361.35,361.35);
\definecolor[named]{drawColor}{rgb}{0.00,0.00,0.00}
\definecolor[named]{fillColor}{rgb}{0.30,0.30,0.30}

\draw[color=drawColor,line cap=round,line join=round,fill=fillColor,] ( 70.27, 60.00) rectangle ( 76.54,123.10);
\definecolor[named]{fillColor}{rgb}{0.50,0.50,0.50}

\draw[color=drawColor,line cap=round,line join=round,fill=fillColor,] ( 76.54, 60.00) rectangle ( 82.80,141.75);
\definecolor[named]{fillColor}{rgb}{0.63,0.63,0.63}

\draw[color=drawColor,line cap=round,line join=round,fill=fillColor,] ( 82.80, 60.00) rectangle ( 89.06,111.86);
\definecolor[named]{fillColor}{rgb}{0.73,0.73,0.73}

\draw[color=drawColor,line cap=round,line join=round,fill=fillColor,] ( 89.06, 60.00) rectangle ( 95.33,121.28);
\definecolor[named]{fillColor}{rgb}{0.82,0.82,0.82}

\draw[color=drawColor,line cap=round,line join=round,fill=fillColor,] ( 95.33, 60.00) rectangle (101.59,118.39);
\definecolor[named]{fillColor}{rgb}{0.90,0.90,0.90}

\draw[color=drawColor,line cap=round,line join=round,fill=fillColor,] (101.59, 60.00) rectangle (107.85,110.33);
\definecolor[named]{fillColor}{rgb}{0.30,0.30,0.30}

\draw[color=drawColor,line cap=round,line join=round,fill=fillColor,] (114.12, 60.00) rectangle (120.38, 91.99);
\definecolor[named]{fillColor}{rgb}{0.50,0.50,0.50}

\draw[color=drawColor,line cap=round,line join=round,fill=fillColor,] (120.38, 60.00) rectangle (126.64, 98.43);
\definecolor[named]{fillColor}{rgb}{0.63,0.63,0.63}

\draw[color=drawColor,line cap=round,line join=round,fill=fillColor,] (126.64, 60.00) rectangle (132.91,104.67);
\definecolor[named]{fillColor}{rgb}{0.73,0.73,0.73}

\draw[color=drawColor,line cap=round,line join=round,fill=fillColor,] (132.91, 60.00) rectangle (139.17, 98.55);
\definecolor[named]{fillColor}{rgb}{0.82,0.82,0.82}

\draw[color=drawColor,line cap=round,line join=round,fill=fillColor,] (139.17, 60.00) rectangle (145.43,100.35);
\definecolor[named]{fillColor}{rgb}{0.90,0.90,0.90}

\draw[color=drawColor,line cap=round,line join=round,fill=fillColor,] (145.43, 60.00) rectangle (151.70, 96.24);
\definecolor[named]{fillColor}{rgb}{0.30,0.30,0.30}

\draw[color=drawColor,line cap=round,line join=round,fill=fillColor,] (157.96, 60.00) rectangle (164.23,152.41);
\definecolor[named]{fillColor}{rgb}{0.50,0.50,0.50}

\draw[color=drawColor,line cap=round,line join=round,fill=fillColor,] (164.23, 60.00) rectangle (170.49,179.33);
\definecolor[named]{fillColor}{rgb}{0.63,0.63,0.63}

\draw[color=drawColor,line cap=round,line join=round,fill=fillColor,] (170.49, 60.00) rectangle (176.75,211.87);
\definecolor[named]{fillColor}{rgb}{0.73,0.73,0.73}

\draw[color=drawColor,line cap=round,line join=round,fill=fillColor,] (176.75, 60.00) rectangle (183.02,122.52);
\definecolor[named]{fillColor}{rgb}{0.82,0.82,0.82}

\draw[color=drawColor,line cap=round,line join=round,fill=fillColor,] (183.02, 60.00) rectangle (189.28,157.54);
\definecolor[named]{fillColor}{rgb}{0.90,0.90,0.90}

\draw[color=drawColor,line cap=round,line join=round,fill=fillColor,] (189.28, 60.00) rectangle (195.54,137.74);
\definecolor[named]{fillColor}{rgb}{0.30,0.30,0.30}

\draw[color=drawColor,line cap=round,line join=round,fill=fillColor,] (201.81, 60.00) rectangle (208.07,279.11);
\definecolor[named]{fillColor}{rgb}{0.50,0.50,0.50}

\draw[color=drawColor,line cap=round,line join=round,fill=fillColor,] (208.07, 60.00) rectangle (214.33,315.14);
\definecolor[named]{fillColor}{rgb}{0.63,0.63,0.63}

\draw[color=drawColor,line cap=round,line join=round,fill=fillColor,] (214.33, 60.00) rectangle (220.60,306.58);
\definecolor[named]{fillColor}{rgb}{0.73,0.73,0.73}

\draw[color=drawColor,line cap=round,line join=round,fill=fillColor,] (220.60, 60.00) rectangle (226.86,257.52);
\definecolor[named]{fillColor}{rgb}{0.82,0.82,0.82}

\draw[color=drawColor,line cap=round,line join=round,fill=fillColor,] (226.86, 60.00) rectangle (233.12,274.30);
\definecolor[named]{fillColor}{rgb}{0.90,0.90,0.90}

\draw[color=drawColor,line cap=round,line join=round,fill=fillColor,] (233.12, 60.00) rectangle (239.39,279.94);
\definecolor[named]{fillColor}{rgb}{0.30,0.30,0.30}

\draw[color=drawColor,line cap=round,line join=round,fill=fillColor,] (245.65, 60.00) rectangle (251.92,275.46);
\definecolor[named]{fillColor}{rgb}{0.50,0.50,0.50}

\draw[color=drawColor,line cap=round,line join=round,fill=fillColor,] (251.92, 60.00) rectangle (258.18,303.20);
\definecolor[named]{fillColor}{rgb}{0.63,0.63,0.63}

\draw[color=drawColor,line cap=round,line join=round,fill=fillColor,] (258.18, 60.00) rectangle (264.44,301.94);
\definecolor[named]{fillColor}{rgb}{0.73,0.73,0.73}

\draw[color=drawColor,line cap=round,line join=round,fill=fillColor,] (264.44, 60.00) rectangle (270.71,251.35);
\definecolor[named]{fillColor}{rgb}{0.82,0.82,0.82}

\draw[color=drawColor,line cap=round,line join=round,fill=fillColor,] (270.71, 60.00) rectangle (276.97,256.07);
\definecolor[named]{fillColor}{rgb}{0.90,0.90,0.90}

\draw[color=drawColor,line cap=round,line join=round,fill=fillColor,] (276.97, 60.00) rectangle (283.23,245.91);
\definecolor[named]{fillColor}{rgb}{0.30,0.30,0.30}

\draw[color=drawColor,line cap=round,line join=round,fill=fillColor,] (289.50, 60.00) rectangle (295.76,292.11);
\definecolor[named]{fillColor}{rgb}{0.50,0.50,0.50}

\draw[color=drawColor,line cap=round,line join=round,fill=fillColor,] (295.76, 60.00) rectangle (302.02,304.17);
\definecolor[named]{fillColor}{rgb}{0.63,0.63,0.63}

\draw[color=drawColor,line cap=round,line join=round,fill=fillColor,] (302.02, 60.00) rectangle (308.29,321.59);
\definecolor[named]{fillColor}{rgb}{0.73,0.73,0.73}

\draw[color=drawColor,line cap=round,line join=round,fill=fillColor,] (308.29, 60.00) rectangle (314.55,293.41);
\definecolor[named]{fillColor}{rgb}{0.82,0.82,0.82}

\draw[color=drawColor,line cap=round,line join=round,fill=fillColor,] (314.55, 60.00) rectangle (320.81,272.78);
\definecolor[named]{fillColor}{rgb}{0.90,0.90,0.90}

\draw[color=drawColor,line cap=round,line join=round,fill=fillColor,] (320.81, 60.00) rectangle (327.08,286.93);
\end{scope}
\begin{scope}
\path[clip] (  0.00,  0.00) rectangle (361.35,361.35);
\definecolor[named]{drawColor}{rgb}{0.00,0.00,0.00}

\node[color=drawColor,anchor=base,inner sep=0pt, outer sep=0pt, scale=  1.00] at ( 89.06, 36.00) {CORI};

\node[color=drawColor,anchor=base,inner sep=0pt, outer sep=0pt, scale=  1.00] at (130.75, 36.00) {BigDocLM};

\node[color=drawColor,anchor=base,inner sep=0pt, outer sep=0pt, scale=  1.00] at (176.75, 36.00) {GAVG};

\node[color=drawColor,anchor=base,inner sep=0pt, outer sep=0pt, scale=  1.00] at (220.60, 36.00) {ReDDE};

\node[color=drawColor,anchor=base,inner sep=0pt, outer sep=0pt, scale=  1.00] at (264.44, 36.00) {CRCS};

\node[color=drawColor,anchor=base,inner sep=0pt, outer sep=0pt, scale=  1.00] at (308.29, 36.00) {SB1};
\end{scope}
\begin{scope}
\path[clip] (  0.00,  0.00) rectangle (361.35,361.35);
\end{scope}
\begin{scope}
\path[clip] ( 60.00, 60.00) rectangle (337.35,325.35);
\definecolor[named]{drawColor}{rgb}{0.00,0.00,0.00}
\definecolor[named]{fillColor}{rgb}{0.30,0.30,0.30}

\draw[color=drawColor,line cap=round,line join=round,fill=fillColor,] ( 69.00,316.35) rectangle ( 76.20,310.35);
\definecolor[named]{fillColor}{rgb}{0.50,0.50,0.50}

\draw[color=drawColor,line cap=round,line join=round,fill=fillColor,] ( 69.00,304.35) rectangle ( 76.20,298.35);
\definecolor[named]{fillColor}{rgb}{0.63,0.63,0.63}

\draw[color=drawColor,line cap=round,line join=round,fill=fillColor,] ( 69.00,292.35) rectangle ( 76.20,286.35);
\definecolor[named]{fillColor}{rgb}{0.73,0.73,0.73}

\draw[color=drawColor,line cap=round,line join=round,fill=fillColor,] ( 69.00,280.35) rectangle ( 76.20,274.35);
\definecolor[named]{fillColor}{rgb}{0.82,0.82,0.82}

\draw[color=drawColor,line cap=round,line join=round,fill=fillColor,] ( 69.00,268.35) rectangle ( 76.20,262.35);
\definecolor[named]{fillColor}{rgb}{0.90,0.90,0.90}

\draw[color=drawColor,line cap=round,line join=round,fill=fillColor,] ( 69.00,256.35) rectangle ( 76.20,250.35);

\node[color=drawColor,anchor=base west,inner sep=0pt, outer sep=0pt, scale=  1.00] at ( 85.20,309.91) {pages-random};

\node[color=drawColor,anchor=base west,inner sep=0pt, outer sep=0pt, scale=  1.00] at ( 85.20,297.91) {pages-top};

\node[color=drawColor,anchor=base west,inner sep=0pt, outer sep=0pt, scale=  1.00] at ( 85.20,285.91) {pages-zipf};

\node[color=drawColor,anchor=base west,inner sep=0pt, outer sep=0pt, scale=  1.00] at ( 85.20,273.91) {snippets-random};

\node[color=drawColor,anchor=base west,inner sep=0pt, outer sep=0pt, scale=  1.00] at ( 85.20,261.91) {snippets-top};

\node[color=drawColor,anchor=base west,inner sep=0pt, outer sep=0pt, scale=  1.00] at ( 85.20,249.91) {snippets-zipf};
\end{scope}
\begin{scope}
\path[clip] (  0.00,  0.00) rectangle (361.35,361.35);
\definecolor[named]{drawColor}{rgb}{0.00,0.00,0.00}

\node[color=drawColor,anchor=base,inner sep=0pt, outer sep=0pt, scale=  1.50] at (198.67, 12.00) {Methods};

\node[rotate= 90.00,color=drawColor,anchor=base,inner sep=0pt, outer sep=0pt, scale=  1.50] at ( 24.00,192.68) {Recall};
\end{scope}
\begin{scope}
\path[clip] (  0.00,  0.00) rectangle (361.35,361.35);
\definecolor[named]{drawColor}{rgb}{0.00,0.00,0.00}

\draw[color=drawColor,line cap=round,line join=round,fill opacity=0.00,] ( 60.00, 60.00) -- ( 60.00,325.35);

\draw[color=drawColor,line cap=round,line join=round,fill opacity=0.00,] ( 60.00, 60.00) -- ( 54.00, 60.00);

\draw[color=drawColor,line cap=round,line join=round,fill opacity=0.00,] ( 60.00,113.07) -- ( 54.00,113.07);

\draw[color=drawColor,line cap=round,line join=round,fill opacity=0.00,] ( 60.00,166.14) -- ( 54.00,166.14);

\draw[color=drawColor,line cap=round,line join=round,fill opacity=0.00,] ( 60.00,219.21) -- ( 54.00,219.21);

\draw[color=drawColor,line cap=round,line join=round,fill opacity=0.00,] ( 60.00,272.28) -- ( 54.00,272.28);

\draw[color=drawColor,line cap=round,line join=round,fill opacity=0.00,] ( 60.00,325.35) -- ( 54.00,325.35);

\node[rotate= 90.00,color=drawColor,anchor=base,inner sep=0pt, outer sep=0pt, scale=  1.00] at ( 48.00, 60.00) {0.0};

\node[rotate= 90.00,color=drawColor,anchor=base,inner sep=0pt, outer sep=0pt, scale=  1.00] at ( 48.00,113.07) {0.1};

\node[rotate= 90.00,color=drawColor,anchor=base,inner sep=0pt, outer sep=0pt, scale=  1.00] at ( 48.00,166.14) {0.2};

\node[rotate= 90.00,color=drawColor,anchor=base,inner sep=0pt, outer sep=0pt, scale=  1.00] at ( 48.00,219.21) {0.3};

\node[rotate= 90.00,color=drawColor,anchor=base,inner sep=0pt, outer sep=0pt, scale=  1.00] at ( 48.00,272.28) {0.4};

\node[rotate= 90.00,color=drawColor,anchor=base,inner sep=0pt, outer sep=0pt, scale=  1.00] at ( 48.00,325.35) {0.5};

\end{scope}
\end{tikzpicture}

}
\caption{Results}
\label{results}
\end{figure}

\noindent \textbf{Methods comparison} \\
\noindent 
In addition we can make the following observations: First, small document models
perform better than big document models, for example when comparing GAVG with CORI and LM.
Second, from the big document models, CORI is consistently better than LM. 
Third, ReDDE is consistently better than CRCS. Fourth, for methods that
take size into account, sampling using pages achieves higher performance. For the other methods,
the results between the different sampling methods are mixed.\\

\noindent \textbf{Performance on non-WSE's} \\
\noindent 
An additional analysis was done by filtering out the 10 general web search engines from the results.
For all methods, performance decreased. However, we found the same observations as just outlined with
the results containing all search engines.\\

\noindent \textbf{Baselines} \\
\noindent 
The results for the baselines (see Section \ref{baselines section}) that do not depend on the sampling are presented in Table \ref{baseline results}.
First, we find that always selecting the 5 major search engines results in a relatively high performance.
This confirms that the queries from the Web track are highly biased towards a general web search scenario.
In addition, a simple ranking based only on size (using the ClueWeb Zipf method) also performs well (indicated as Size Based 2).
Using only Google, the average recall is 0.574 and the precision is 0.392 for the top 10 results.\\

\begin{table}[h]
\centering
\begin{tabular}{lll}
\hline \textbf{Baseline} & \textbf{Recall at 5} & \textbf{Precision at 5} \\
\hline Popular & 0.646  & 0.2144 \\ 
 Size Based 2 &  0.477 & 0.1588 \\ 
\hline
\end{tabular}
\caption{Baseline results - Better than Rel}
\label{baseline results}
\end{table}

Overall, the performance of the resource selection methods in combination with the used query samplings
is relatively low. In the next section we analyze the effect of size and the limited samplings in more depth.

\subsection{Discussion}
\subsubsection{Importance of size}

We found that the ranking methods that incorporate size perform relatively well. In this section
we provide further analyses on the effect of size.  
First we ask:

\begin{quote}
To what extent do the resource selection methods correlate with a size based ranking? 
\end{quote}

We calculate the correlations between the ranking of the resource selection methods
and a sized based ranking (Table \ref{correlation size}) using Kendalls Tau. Because only a partial set of collections is ranked, we only calculate 
the correlation over the resources that are considered by the ranking method. We report the average correlation of a method
by averaging across queries and sampling methods. We find that the methods that take size explicitly into account have a high correlation,
while the other methods have a small negative correlation.

\begin{table}[ht]
\centering
\begin{tabular}{ll}
\hline  \textbf{Method} & \textbf{$\tau$} \\
\hline  Big document LM &  -0.085\\
 CORI &  -0.017\\
 GAVG &  -0.013\\
 ReDDE & 0.595 \\
 CRCS &  0.487\\
\hline
\end{tabular}
\caption{Correlation with size based ranking}
\label{correlation size}
\end{table}

To illustrate why methods such as ReDDE have such a high correlation with size,
we provide an example calculation of the scaling factor that is used: $\frac{|C_i|}{|S_i|} $.
For a resource like Google, the ClueWeb method estimated a size of $|C_i| = 561,965,753$.
The sampling size for \emph{Zipf} samples using pages was $|S_i| = 389$, resulting in a scaling factor of 
561,965,753/389 = 1,444,642. Compare this for example to Fox Sports, which gets a scaling factor of  only 16,027/139 = 115.
It is questionable whether
such a high influence of size is desired.

Next, we investigate the following question:

\begin{quote}
How is performance influenced by the choice of size estimation method? 
\end{quote}

Table \ref{size estimation methods table} shows
a comparison of the used methods by measuring the recall and precision at 5.
Both methods perform similarly, although using \emph{top} samples results in a low performance,
probably because these queries are far from random.

\begin{table}[ht]
\centering

\begin{tabular}{lll}
\hline  \textbf{Method} & \textbf{Recall at 5} & \textbf{Precision at 5}\\
\hline Clueweb I  - Zipf &   0.493 &  0.1704\\
Clueweb II - Zipf & 0.467 & 0.1856\\
 QueryPool - Top &  0.117 & 0.0644 \\
 QueryPool - Zipf &  0.448 & 0.1932 \\
\hline
\end{tabular}
\caption{SB1 - Zipf pages sampling}
\label{size estimation methods table}
\end{table}

Table \ref{size estimation methods - major removed} shows the results using only non-WSE's.
\begin{table}[ht]
\centering
\begin{tabular}{lll}
\hline  \textbf{Method} & \textbf{Recall at 5} & \textbf{Precision at 5}\\
\hline Clueweb I  - Zipf &   0.150 & 0.091 \\
Clueweb II - Zipf &  0.172  & 0.105 \\
 QueryPool - Top &   0.110 &  0.067\\
 QueryPool - Zipf &  0.184  &  0.114\\
\hline
\end{tabular}
\caption{SB1 - Zipf pages sampling - non-WSE}
\label{size estimation methods - major removed}
\end{table}

One striking observation of the results discussed above is that the size based rankings
perform so well. We will now look closer into this observation and we ask:

\begin{quote}
When is a purely size based ranking effective, and when are more sophisticated
resource selection methods effective?
\end{quote}

We make the following hypothesis: on queries for which the ranking quality of the documents
in the sample index is low (for example when the sample index has not enough samples matching
 the query), a ranking only based on size will perform better than resource sampling methods. 
 
In our analysis we will analyze  the SB1 ranking using size estimated by the ClueWeb size estimation method
based on page-based Zipf sampling. We will compare with the other resource selection methods,
using an index based on Zipf sampling pages.

We make pairs of the SB1 ranking with the other resource selection methods,
and calculate the difference between recall@5. A positive difference indicates that the SB1 ranking performed better than
the other method.

Next, we calculate the correlation between the difference and an estimate of the ranking quality
for the queries. We use the Simplified Clarity Score (SCS) as proposed in He and Ounis \cite{springerlink:10.1007/978-3-540-30213-1_5}, which is based on query clarity \cite{Cronen-Townsend:2002:PQP:564376.564429} . He and Ounis \cite{springerlink:10.1007/978-3-540-30213-1_5} found a strong positive correlation between SCS
and AP for short queries. A higher value for SCS therefore corresponds with a better ranking quality.

\[SCS = \sum_Q P_{ml}(w|Q) log_2 \frac{P_{ml}(w|Q) }{P_{coll}(w)}\]

with

\[P_{ml}(w|Q)  = \frac{qtf}{ql}\]

Where $qtf$ is the query term frequency, and $ql$ is the query length. 
$P_{coll}(w)$ is the probability of term w in the collection.
The found correlations are shown in Table \ref{correlation scs}.

\begin{table}[ht]
\centering
\begin{tabular}{ll}
\hline  \textbf{Method} & \textbf{Correlation} \\
\hline  Big document LM & -0.140 \\
 CORI & -0.153 \\
 GAVG &  -0.067\\
 ReDDE &  -0.142\\
 CRCS &  -0.128\\
\hline
\end{tabular}
\caption{Pearson correlation with SCS and performance of SB1 compared with other methods}
\label{correlation scs}
\end{table}

For every method, we find a low but consistently negative correlation. This shows that the SB1 method has a tendency to be more effective compared to
other methods for queries where the ranking quality is (predicted to be)
low. We therefore expect a possible future strategy for resource selection methods that have to deal with sparse samples is to back off to a size based ranking for queries that are considered difficult.

\subsubsection{Is sampling enough?}
In this section we analyze how performance is influenced by the amount of samplings.\\

\noindent \textbf{How big is the ranking problem?} \\
\noindent 
How many resource match at least one of the query terms such that it is considered in the ranking?
The results
are presented in Table \ref{average to rank} (avg of total). On average, the number of resources to rank is around 50-60 out of 108,
and using pages the number is consistently higher than using snippets.\\

\begin{table}
\centering
\begin{tabular}{lll}
\hline \textbf{Sampling method} & \textbf{Avg of total}  & \textbf{Avg of relevant} \\
\hline Zipf-pages & 0.51 & 0.72\\ 
 Zipf-snippets & 0.47 &  0.61 \\ 
 Top-pages & 0.55 & 0.76\\ 
 Top-snippets & 0.46 &  0.57\\ 
 Random-pages & 0.58  &  0.73\\ 
 Random-snippets &  0.50 &  0.59\\ 
\hline
\end{tabular}
\caption{Average fraction of resources considered for Web TREC 2010 queries}
\label{average to rank}
\end{table}

\noindent \textbf{How many resources are we missing?} \\
Next we analyze how many resources we are not even considering for ranking because
of the limited sampling. In particular, for each query we divide the number of relevant resources
(a resource that has at least one relevant document)  that
match at least one of the query terms
by the total number of relevant resources for that query.
We then average across queries. Results are found in Table \ref{average to rank} (avg of relevant).
For example, for a given query 60 resources are considered for ranking, but only 20 of them are
relevant. If the total number of relevant resources is 50, it means that we only consider 20/50 = 0.4
of the relevant resources for ranking. We find that on average only 50-75\% of the relevant resources
are considered for ranking, indicating that many relevant resources are missed because of the limited samplings.\\

\noindent \textbf{Varying the amount of samples} \\
\noindent Now, we are interested in how the amount of sampling influences performance.
We vary the amount of sampling available, by taking proportions of the obtained snippet samples
by randomly selecting 0.25, 0.50 and 0.75 of the queries. Results are plotted in Figure \ref{sampling size}. We analyze the performance
of  two methods ReDDE and GAVG.
We observe that ReDDE shows an increasing performance
as more data is available. The trend with GAVG is more noisy, although
the pattern still suggests an increasing line.
\begin{figure}[htb]
\center
\subfigure[ReDDE]{
\scalebox{0.30}{

\begin{tikzpicture}[x=1pt,y=1pt]
\definecolor[named]{drawColor}{rgb}{0.00,0.00,0.00}
\definecolor[named]{fillColor}{rgb}{1.00,1.00,1.00}
\fill[color=fillColor,fill opacity=0.00,] (0,0) rectangle (361.35,361.35);
\begin{scope}
\path[clip] ( 84.00, 60.00) rectangle (337.35,313.35);
\definecolor[named]{drawColor}{rgb}{0.00,0.00,1.00}

\draw[color=drawColor,line cap=round,line join=round,fill opacity=0.00,] ( 93.38,136.86) --
	(171.58,177.95) --
	(249.77,199.68) --
	(327.97,204.03);

\draw[color=drawColor,line cap=round,line join=round,fill opacity=0.00,] ( 93.38,136.86) circle (  2.25);

\draw[color=drawColor,line cap=round,line join=round,fill opacity=0.00,] (171.58,177.95) circle (  2.25);

\draw[color=drawColor,line cap=round,line join=round,fill opacity=0.00,] (249.77,199.68) circle (  2.25);

\draw[color=drawColor,line cap=round,line join=round,fill opacity=0.00,] (327.97,204.03) circle (  2.25);
\end{scope}
\begin{scope}
\path[clip] (  0.00,  0.00) rectangle (361.35,361.35);
\end{scope}
\begin{scope}
\path[clip] ( 84.00, 60.00) rectangle (337.35,313.35);
\definecolor[named]{drawColor}{rgb}{0.00,1.00,0.00}

\draw[color=drawColor,line cap=round,line join=round,fill opacity=0.00,] ( 93.38,157.99) --
	(171.58,179.27) --
	(249.77,236.65) --
	(327.97,228.75);

\draw[color=drawColor,line cap=round,line join=round,fill opacity=0.00,] ( 93.38,157.99) circle (  2.25);

\draw[color=drawColor,line cap=round,line join=round,fill opacity=0.00,] (171.58,179.27) circle (  2.25);

\draw[color=drawColor,line cap=round,line join=round,fill opacity=0.00,] (249.77,236.65) circle (  2.25);

\draw[color=drawColor,line cap=round,line join=round,fill opacity=0.00,] (327.97,228.75) circle (  2.25);
\definecolor[named]{drawColor}{rgb}{1.00,0.00,0.00}

\draw[color=drawColor,line cap=round,line join=round,fill opacity=0.00,] ( 93.38,168.64) --
	(171.58,211.69) --
	(249.77,209.45) --
	(327.97,237.05);

\draw[color=drawColor,line cap=round,line join=round,fill opacity=0.00,] ( 93.38,168.64) circle (  2.25);

\draw[color=drawColor,line cap=round,line join=round,fill opacity=0.00,] (171.58,211.69) circle (  2.25);

\draw[color=drawColor,line cap=round,line join=round,fill opacity=0.00,] (249.77,209.45) circle (  2.25);

\draw[color=drawColor,line cap=round,line join=round,fill opacity=0.00,] (327.97,237.05) circle (  2.25);
\end{scope}
\begin{scope}
\path[clip] (  0.00,  0.00) rectangle (361.35,361.35);
\definecolor[named]{drawColor}{rgb}{0.00,0.00,0.00}

\draw[color=drawColor,line cap=round,line join=round,fill opacity=0.00,] ( 84.00, 69.38) -- ( 84.00,303.97);

\draw[color=drawColor,line cap=round,line join=round,fill opacity=0.00,] ( 84.00, 69.38) -- ( 78.00, 69.38);

\draw[color=drawColor,line cap=round,line join=round,fill opacity=0.00,] ( 84.00,186.67) -- ( 78.00,186.67);

\draw[color=drawColor,line cap=round,line join=round,fill opacity=0.00,] ( 84.00,303.97) -- ( 78.00,303.97);

\node[rotate= 90.00,color=drawColor,anchor=base,inner sep=0pt, outer sep=0pt, scale=  3.00] at ( 72.00, 69.38) {0.20};

\node[rotate= 90.00,color=drawColor,anchor=base,inner sep=0pt, outer sep=0pt, scale=  3.00] at ( 72.00,186.67) {0.35};

\node[rotate= 90.00,color=drawColor,anchor=base,inner sep=0pt, outer sep=0pt, scale=  3.00] at ( 72.00,303.97) {0.50};

\draw[color=drawColor,line cap=round,line join=round,fill opacity=0.00,] ( 93.38, 60.00) -- (327.97, 60.00);

\draw[color=drawColor,line cap=round,line join=round,fill opacity=0.00,] ( 93.38, 60.00) -- ( 93.38, 54.00);

\draw[color=drawColor,line cap=round,line join=round,fill opacity=0.00,] (327.97, 60.00) -- (327.97, 54.00);

\node[color=drawColor,anchor=base,inner sep=0pt, outer sep=0pt, scale=  3.00] at ( 93.38, 36.00) {0.25};

\node[color=drawColor,anchor=base,inner sep=0pt, outer sep=0pt, scale=  3.00] at (327.97, 36.00) {1.00};
\end{scope}
\begin{scope}
\path[clip] ( 84.00, 60.00) rectangle (337.35,313.35);
\definecolor[named]{drawColor}{rgb}{0.00,0.00,1.00}

\draw[color=drawColor,line width= 1.0pt,line cap=round,line join=round,fill opacity=0.00,] (128.61,296.02) -- (173.61,296.02);
\definecolor[named]{drawColor}{rgb}{0.00,1.00,0.00}

\draw[color=drawColor,line width= 1.0pt,line cap=round,line join=round,fill opacity=0.00,] (128.61,266.02) -- (173.61,266.02);
\definecolor[named]{drawColor}{rgb}{1.00,0.00,0.00}

\draw[color=drawColor,line width= 1.0pt,line cap=round,line join=round,fill opacity=0.00,] (128.61,236.02) -- (173.61,236.02);
\definecolor[named]{drawColor}{rgb}{0.00,0.00,0.00}

\node[color=drawColor,anchor=base west,inner sep=0pt, outer sep=0pt, scale=  2.50] at (196.11,287.41) {Random};

\node[color=drawColor,anchor=base west,inner sep=0pt, outer sep=0pt, scale=  2.50] at (196.11,257.41) {Top};

\node[color=drawColor,anchor=base west,inner sep=0pt, outer sep=0pt, scale=  2.50] at (196.11,227.41) {Zipf};
\end{scope}
\begin{scope}
\path[clip] (  0.00,  0.00) rectangle (361.35,361.35);
\definecolor[named]{drawColor}{rgb}{0.00,0.00,0.00}

\node[color=drawColor,anchor=base,inner sep=0pt, outer sep=0pt, scale=  3.00] at (210.67, 12.00) {Sampling percentage};

\node[rotate= 90.00,color=drawColor,anchor=base,inner sep=0pt, outer sep=0pt, scale=  3.00] at ( 48.00,186.67) {Recall at 5};
\end{scope}
\end{tikzpicture}

}
}
\hspace{-5mm}
\subfigure[GAVG]{
\scalebox{0.30}{

\begin{tikzpicture}[x=1pt,y=1pt]
\definecolor[named]{drawColor}{rgb}{0.00,0.00,0.00}
\definecolor[named]{fillColor}{rgb}{1.00,1.00,1.00}
\fill[color=fillColor,fill opacity=0.00,] (0,0) rectangle (361.35,361.35);
\begin{scope}
\path[clip] ( 84.00, 60.00) rectangle (337.35,313.35);
\definecolor[named]{drawColor}{rgb}{0.00,0.00,1.00}

\draw[color=drawColor,line cap=round,line join=round,fill opacity=0.00,] ( 93.38, 77.29) --
	(171.58, 94.47) --
	(249.77, 71.31) --
	(327.97,111.15);

\draw[color=drawColor,line cap=round,line join=round,fill opacity=0.00,] ( 93.38, 77.29) circle (  2.25);

\draw[color=drawColor,line cap=round,line join=round,fill opacity=0.00,] (171.58, 94.47) circle (  2.25);

\draw[color=drawColor,line cap=round,line join=round,fill opacity=0.00,] (249.77, 71.31) circle (  2.25);

\draw[color=drawColor,line cap=round,line join=round,fill opacity=0.00,] (327.97,111.15) circle (  2.25);
\end{scope}
\begin{scope}
\path[clip] (  0.00,  0.00) rectangle (361.35,361.35);
\end{scope}
\begin{scope}
\path[clip] ( 84.00, 60.00) rectangle (337.35,313.35);
\definecolor[named]{drawColor}{rgb}{0.00,1.00,0.00}

\draw[color=drawColor,line cap=round,line join=round,fill opacity=0.00,] ( 93.38,168.31) --
	(171.58,206.55) --
	(249.77,212.57) --
	(327.97,265.97);

\draw[color=drawColor,line cap=round,line join=round,fill opacity=0.00,] ( 93.38,168.31) circle (  2.25);

\draw[color=drawColor,line cap=round,line join=round,fill opacity=0.00,] (171.58,206.55) circle (  2.25);

\draw[color=drawColor,line cap=round,line join=round,fill opacity=0.00,] (249.77,212.57) circle (  2.25);

\draw[color=drawColor,line cap=round,line join=round,fill opacity=0.00,] (327.97,265.97) circle (  2.25);
\definecolor[named]{drawColor}{rgb}{1.00,0.00,0.00}

\draw[color=drawColor,line cap=round,line join=round,fill opacity=0.00,] ( 93.38,113.33) --
	(171.58,185.48) --
	(249.77,131.18) --
	(327.97,178.41);

\draw[color=drawColor,line cap=round,line join=round,fill opacity=0.00,] ( 93.38,113.33) circle (  2.25);

\draw[color=drawColor,line cap=round,line join=round,fill opacity=0.00,] (171.58,185.48) circle (  2.25);

\draw[color=drawColor,line cap=round,line join=round,fill opacity=0.00,] (249.77,131.18) circle (  2.25);

\draw[color=drawColor,line cap=round,line join=round,fill opacity=0.00,] (327.97,178.41) circle (  2.25);
\end{scope}
\begin{scope}
\path[clip] (  0.00,  0.00) rectangle (361.35,361.35);
\definecolor[named]{drawColor}{rgb}{0.00,0.00,0.00}

\draw[color=drawColor,line cap=round,line join=round,fill opacity=0.00,] ( 84.00, 69.38) -- ( 84.00,303.97);

\draw[color=drawColor,line cap=round,line join=round,fill opacity=0.00,] ( 84.00, 69.38) -- ( 78.00, 69.38);

\draw[color=drawColor,line cap=round,line join=round,fill opacity=0.00,] ( 84.00,186.67) -- ( 78.00,186.67);

\draw[color=drawColor,line cap=round,line join=round,fill opacity=0.00,] ( 84.00,303.97) -- ( 78.00,303.97);

\node[rotate= 90.00,color=drawColor,anchor=base,inner sep=0pt, outer sep=0pt, scale=  3.00] at ( 72.00, 69.38) {0.10};

\node[rotate= 90.00,color=drawColor,anchor=base,inner sep=0pt, outer sep=0pt, scale=  3.00] at ( 72.00,186.67) {0.15};

\node[rotate= 90.00,color=drawColor,anchor=base,inner sep=0pt, outer sep=0pt, scale=  3.00] at ( 72.00,303.97) {0.20};

\draw[color=drawColor,line cap=round,line join=round,fill opacity=0.00,] ( 93.38, 60.00) -- (327.97, 60.00);

\draw[color=drawColor,line cap=round,line join=round,fill opacity=0.00,] ( 93.38, 60.00) -- ( 93.38, 54.00);

\draw[color=drawColor,line cap=round,line join=round,fill opacity=0.00,] (327.97, 60.00) -- (327.97, 54.00);

\node[color=drawColor,anchor=base,inner sep=0pt, outer sep=0pt, scale=  3.00] at ( 93.38, 36.00) {0.25};

\node[color=drawColor,anchor=base,inner sep=0pt, outer sep=0pt, scale=  3.00] at (327.97, 36.00) {1.00};
\end{scope}
\begin{scope}
\path[clip] ( 84.00, 60.00) rectangle (337.35,313.35);
\definecolor[named]{drawColor}{rgb}{0.00,0.00,1.00}

\draw[color=drawColor,line width= 1.0pt,line cap=round,line join=round,fill opacity=0.00,] (141.27,283.35) -- (186.27,283.35);
\definecolor[named]{drawColor}{rgb}{0.00,1.00,0.00}

\draw[color=drawColor,line width= 1.0pt,line cap=round,line join=round,fill opacity=0.00,] (141.27,253.35) -- (186.27,253.35);
\definecolor[named]{drawColor}{rgb}{1.00,0.00,0.00}

\draw[color=drawColor,line width= 1.0pt,line cap=round,line join=round,fill opacity=0.00,] (141.27,223.35) -- (186.27,223.35);
\definecolor[named]{drawColor}{rgb}{0.00,0.00,0.00}

\node[color=drawColor,anchor=base west,inner sep=0pt, outer sep=0pt, scale=  2.50] at (208.77,274.74) {Random};

\node[color=drawColor,anchor=base west,inner sep=0pt, outer sep=0pt, scale=  2.50] at (208.77,244.74) {Top};

\node[color=drawColor,anchor=base west,inner sep=0pt, outer sep=0pt, scale=  2.50] at (208.77,214.74) {Zipf};
\end{scope}
\begin{scope}
\path[clip] (  0.00,  0.00) rectangle (361.35,361.35);
\definecolor[named]{drawColor}{rgb}{0.00,0.00,0.00}

\node[color=drawColor,anchor=base,inner sep=0pt, outer sep=0pt, scale=  3.00] at (210.67, 12.00) {Sampling percentage};

\node[rotate= 90.00,color=drawColor,anchor=base,inner sep=0pt, outer sep=0pt, scale=  3.00] at ( 48.00,186.67) {Recall at 5};
\end{scope}
\end{tikzpicture}

}
}
\caption{Effect of sampling size}
\label{sampling size}
\end{figure}
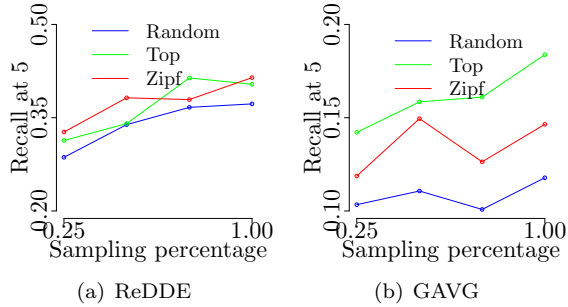

Overall, we can conclude that the limited amount of sampling did have a large
influence on the performance. We expect, and suggest for future work, that more
samplings are desirable for this setting.


\section{Dataset robustness}
We will use the experiments from the previous section
to assess the robustness of the test collection for research in resource selection for federated search.
We will experiment with two measures, the Cronbach Alpha  coefficient and the correlation between rankings
based on subsets.

\subsection{Cronbach Alpha}
The Cronbach Alpha is a coefficient for measuring the reliability of tests. Recently it has been used to measure the reliability
of IR datasets (e.g. \cite{Bodoff:2007:TTA:1277741.1277805}, \cite{Harpale:2010:CNM:1871437.1871509}).
The coefficient gives an indication to what degree the questions test a single construct.
The coefficient will be high, when there is a high correlation between the queries as well as with
the end score (for example MAP or average recall@5). 
The coefficient is calculated as follows: 

\[\hat{\alpha} = \frac{k}{k-1} (1 - \frac{\sum_i \hat{\sigma}_i^2}{\hat{\sigma}_x^2})\]

Where $k$ is the number of queries, $\hat{\sigma}_i^2$} is the estimated
variance for query $i$ and $\hat{\sigma}_x^2$ is the estimated variance
of the total scores. The test is simple to calculate, but it does not take into account
individual variance of items. 
The test requires as input existing runs, we use the runs performed for our empirical comparison of the resource selection methods
and use as scores the recall@5.
We find a value of 0.98, which indicates a very high degree
of reliability.

\subsection{Subset topics}
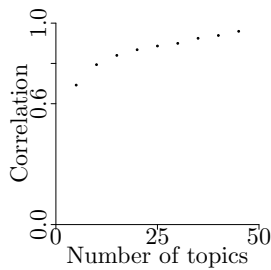
\begin{figure}[htb]
\center
\scalebox{0.30}{

\begin{tikzpicture}[x=1pt,y=1pt]
\definecolor[named]{drawColor}{rgb}{0.00,0.00,0.00}
\definecolor[named]{fillColor}{rgb}{1.00,1.00,1.00}
\fill[color=fillColor,fill opacity=0.00,] (0,0) rectangle (361.35,361.35);
\begin{scope}
\path[clip] ( 84.00, 60.00) rectangle (337.35,313.35);
\definecolor[named]{drawColor}{rgb}{0.00,0.00,0.00}
\definecolor[named]{fillColor}{rgb}{0.00,0.00,0.00}

\draw[color=drawColor,line cap=round,line join=round,fill=fillColor,] (109.33,235.42) circle (  1.12);

\draw[color=drawColor,line cap=round,line join=round,fill=fillColor,] (134.67,261.13) circle (  1.12);

\draw[color=drawColor,line cap=round,line join=round,fill=fillColor,] (160.00,272.81) circle (  1.12);

\draw[color=drawColor,line cap=round,line join=round,fill=fillColor,] (185.34,279.92) circle (  1.12);

\draw[color=drawColor,line cap=round,line join=round,fill=fillColor,] (210.67,284.59) circle (  1.12);

\draw[color=drawColor,line cap=round,line join=round,fill=fillColor,] (236.01,287.93) circle (  1.12);

\draw[color=drawColor,line cap=round,line join=round,fill=fillColor,] (261.34,294.29) circle (  1.12);

\draw[color=drawColor,line cap=round,line join=round,fill=fillColor,] (286.68,297.85) circle (  1.12);

\draw[color=drawColor,line cap=round,line join=round,fill=fillColor,] (312.01,303.07) circle (  1.12);
\end{scope}
\begin{scope}
\path[clip] (  0.00,  0.00) rectangle (361.35,361.35);
\end{scope}
\begin{scope}
\path[clip] (  0.00,  0.00) rectangle (361.35,361.35);
\definecolor[named]{drawColor}{rgb}{0.00,0.00,0.00}

\draw[color=drawColor,line cap=round,line join=round,fill opacity=0.00,] ( 84.00, 60.00) -- ( 84.00,313.35);

\draw[color=drawColor,line cap=round,line join=round,fill opacity=0.00,] ( 84.00, 60.00) -- ( 78.00, 60.00);

\draw[color=drawColor,line cap=round,line join=round,fill opacity=0.00,] ( 84.00,212.01) -- ( 78.00,212.01);

\draw[color=drawColor,line cap=round,line join=round,fill opacity=0.00,] ( 84.00,262.68) -- ( 78.00,262.68);

\draw[color=drawColor,line cap=round,line join=round,fill opacity=0.00,] ( 84.00,313.35) -- ( 78.00,313.35);

\node[rotate= 90.00,color=drawColor,anchor=base,inner sep=0pt, outer sep=0pt, scale=  3.00] at ( 72.00, 60.00) {0.0};

\node[rotate= 90.00,color=drawColor,anchor=base,inner sep=0pt, outer sep=0pt, scale=  3.00] at ( 72.00,212.01) {0.6};

\node[rotate= 90.00,color=drawColor,anchor=base,inner sep=0pt, outer sep=0pt, scale=  3.00] at ( 72.00,313.35) {1.0};

\draw[color=drawColor,line cap=round,line join=round,fill opacity=0.00,] ( 84.00, 60.00) -- (337.35, 60.00);

\draw[color=drawColor,line cap=round,line join=round,fill opacity=0.00,] ( 84.00, 60.00) -- ( 84.00, 54.00);

\draw[color=drawColor,line cap=round,line join=round,fill opacity=0.00,] (210.67, 60.00) -- (210.67, 54.00);

\draw[color=drawColor,line cap=round,line join=round,fill opacity=0.00,] (337.35, 60.00) -- (337.35, 54.00);

\node[color=drawColor,anchor=base,inner sep=0pt, outer sep=0pt, scale=  3.00] at ( 84.00, 36.00) {0};

\node[color=drawColor,anchor=base,inner sep=0pt, outer sep=0pt, scale=  3.00] at (210.67, 36.00) {25};

\node[color=drawColor,anchor=base,inner sep=0pt, outer sep=0pt, scale=  3.00] at (337.35, 36.00) {50};
\end{scope}
\begin{scope}
\path[clip] (  0.00,  0.00) rectangle (361.35,361.35);
\definecolor[named]{drawColor}{rgb}{0.00,0.00,0.00}

\node[color=drawColor,anchor=base,inner sep=0pt, outer sep=0pt, scale=  3.00] at (210.67, 12.00) {Number of topics};

\node[rotate= 90.00,color=drawColor,anchor=base,inner sep=0pt, outer sep=0pt, scale=  3.00] at ( 48.00,186.67) {Correlation};
\end{scope}
\end{tikzpicture}

}
\caption{Correlation with full ranking}
\label{correlationsubset}
\end{figure}

Next we analyze how many topics are needed to obtain a good ranking of the systems.
We select random subsets from the topics, starting at size 5 and increasing with step size 5 to
the full set of topics. For each subset size, we take multiple random samples (10) to calculate an average based on recall at 5.
We calculate the correlation using Kendalls Tau between the ranking obtained on the subset with the ranking
using the full dataset. 
The found correlations are shown in Figure \ref{correlationsubset}.

 Around 30 topics, a correlation higher or equal to 0.75 is obtained. An correlation higher
than 0.86 is achieved after 40 topics or more.
 
\section{Conclusions}
A dataset for federated search on the web, reflecting a real web environment, has long been absent.
Recently, a new test collection was released \cite{nguyencikm2012}, containing the results
from more than a hundred real search engines, ranging from large general web search engines
such as Google and Bing, to small domain-specific engines.
In this paper analyses and experiments on resource selection were presented using this new dataset. 

We demonstrated that federated search has a lot of potential. Without the large general web search engines,
an optimal resource selection algorithm would be able to achieve comparable or better performance when looking at the unique
number of relevant results. However, general web search engines are still useful for delivering highly relevant pages
for example to answer navigational queries.

We experimented with several methods to estimate the size of uncooperative resources on the web.
We used a method based on the number of overlapping documents and also introduced a method that uses 
an external reference corpus (ClueWeb09) to estimate the sizes. The sizes were found
to be very effective for the resource selection methods.

Existing resource selection methods were found not to be readily suitable for the web setting.
We found that due to the extremely skewed size distribution of the dataset, the size
of the collections had an enormous influence on the performance. 
Resource selection methods that take the estimated size of a collection into account performed well compared
to methods that do not include size in their ranking. Also, methods that take the size of a collection into account correlate
highly with just a size based ranking. We found that size based rankings tend to be effective
 compared to more sophisticated methods when the (predicted) ranking quality in 
the sample index is low. 

We suggest several future research directions for resource selection  for the web. 
First, is the large influence of size in current resource selection methods
desirable when the sizes are extremely skewed? Second, more research on size estimation is encouraged, especially in combination
with query based sampling methods.
And third, how to handle sparse resource descriptions and how many samples are desirable? Results suggest that sampling more documents is desirable for obtaining a more complete sample based index
and the typical number of 300-500 documents is not sufficient for this setting to obtain a high performance. 


\section{Acknowledgements}
This research was supported by the Dutch national  program  COMMIT  
and the Folktales as Classifiable Texts (FACT) project, which is part of the CATCH
programme funded by the Netherlands Organisation for
Scientific Research (NWO).

 \bibliographystyle{abbrv}


\end{document}